\documentclass[11pt]{article}

\usepackage{aas_macros,amsmath,amssymb,comment,cite,esint,graphicx,mathtools}
\usepackage[margin=.8in,letterpaper]{geometry}
\usepackage[colorlinks=true]{hyperref}
\usepackage[affil-it]{authblk}
\usepackage{subcaption}
\usepackage[utf8]{inputenc}
\usepackage{mathrsfs}
\usepackage{appendix}
\usepackage{amssymb}
\usepackage{float}
\usepackage{color}
\usepackage{cite}
\usepackage{hyperref}
\hypersetup{pageanchor=false}
\usepackage{indentfirst}
\usepackage{url}
\usepackage{float}
\usepackage{caption}
\usepackage[numbers,square,comma,sort&compress,merge]{natbib}
\usepackage{esint}
\usepackage{overpic}
\usepackage{graphicx}
\usepackage{epsf,amsmath,bbold,amsfonts,stmaryrd}
\usepackage{textcomp}
\usepackage{ulem}
\usepackage{tikz}

\numberwithin{equation}{section}
\let\originalleft\left
\let\originalright\right
\renewcommand{\left}{\mathopen{}\mathclose\bgroup\originalleft}
\renewcommand{\right}{\aftergroup\egroup\originalright}
\mathcode`\*="8000
{\catcode`\*=\active\gdef*{\mathclose{}\,\mathopen{}}}

\newcommand{\be}{\begin{equation}}
\newcommand{\ee}{\end{equation}}
\newcommand{\bea}{\setlength\arraycolsep{2pt} \begin{eqnarray}}
\newcommand{\eea}{\end{eqnarray}}

\begin{document}
\title{Energy extraction from a rotating black hole via magnetic reconnection: parameters in reconnection models}

\author{
Ye Shen$^{1}$\thanks{E-mail: shenye199594@stu.pku.edu.cn},~
Ho-Yun YuChih$^{1}$\thanks{E-mail: hyyc@stu.pku.edu.cn; Co-first author}
}
\date{}
\maketitle

\vspace{-10mm}

\begin{center}
	{\it	
      $^1$School of Physics, Peking University, No.5 Yiheyuan Rd, Beijing
		100871, P.R. China\\\vspace{4mm}

	}
\end{center}

\vspace{8mm}

\begin{abstract}
	\vspace{5mm}
Works on the energy extraction from a rotating black hole via magnetic reconnection attract more attentions in recent years. Discussions on this topic, however, are often based on many simplifications, such as assuming a circularly flowing bulk plasma and a fixed orientation angle. A significant gap remains between theoretical models and the magnetic reconnection occurring in real astrophysical scenarios. In our previous work, we investigated the influence of orientation angle on  energy extraction and figured out the differences between the plunging and circularly flowing bulk plasma. We introduced the concept of covering factor to quantify the capability of an accretion system in extracting energy via magnetic reconnection from a rotating black hole. In this work, as an improvement, we extend our discussions by treating the parameters in reconnection models as free parameters, bringing the theoretical model closer to the situations in real astrophysical systems. We employ two reconnection models, in which the geometric index and the guide field fraction are respectively induced. We separately present the dependences of energy extraction on the geometric index and the guide field fraction. More importantly, we propose to define the averaged covering factor weighted by reconnection rate to quantify the overall capability of an accretion system in extracting energy via magnetic reconnection from a rotating black hole, under the assumption that the magnetic reconnection process occurs stochastically and randomly within the ergosphere.
\end{abstract}

\maketitle

\newpage
\baselineskip 18pt

\section{Introduction}
\label{sec:intro}

It is now widely known that a substantial fraction of energy could be extracted from a rotating black hole. Shown by Christodoulou for the first time, only a portion of the black hole mass is irreducible \cite{Christodoulou:1970wf}. For a Kerr black hole with mass $M$ and dimensionless spin parameter $a$, the maximal amount of energy that could be extracted obeys:
\begin{equation}
    E_{\rm ext}=\left[1-\sqrt{\frac{1}{2}\left(1+\sqrt{1-a^2}\right)}\right]Mc^2
\end{equation}
For an extremely spinning black hole ($a\rightarrow 1$), it gives $E_{\rm ext}\simeq 0.29 Mc^2$. Penrose envisioned a thought experiment which provided an avenue to extract energy from a rotating black hole \cite{Penrose}. However, the so-called Penrose process lacks a clear initiation mechanism and was believed to be difficult to actualize in astrophysical scenario \cite{Wald:1974kya,Bardeen:1972fi}. This motivated the propositions of more practical mechanisms, such as the superradiance of a massive bosonic field \cite{Brito:2015oca} and the Blandford-Znajek (BZ) mechanism \cite{BZ}, to achieve energy extraction. Among these, magnetic reconnection, which is believed to occur frequently in the accretion flow around black holes \cite{Ripperda2020,Ripperda:2021zpn,Yuan2024-1,Yuan2024-2}, attracts more attentions in recent years.

As an exploratory study, Koide and Arai analyzed the feasibility of extracting energy via magnetic reconnection from a Kerr black hole \cite{KA2008}. In their work, the ejected plasma was treated as relativistic adiabatic incompressible ball, the magnetic reconnection is treated as a slowly diffusive process and the magnetic field is set to be purely toroidally. Progressively, Comisso and Asenjo creatively investigated the dynamics of expelled plasmoid \cite{CA2021}, produced by fast magnetic reconnection within magnetized bulk plasma \cite{Sironi:2022hnw,French:2022zfv}. They found that the efficiency of energy extraction from a rotating black hole via magnetic reconnection is even higher than the efficiency via BZ mechanism. Energy extraction via magnetic reconnection, via the so called Comisso-Asenjo process specifically, from a rotating black hole has thus been further explored in various spacetimes \cite{Carleo:2022qlv,Wei:2022jbi,Liu:2022qnr,Wang:2022qmg,Zhang:2024ptp,Rodriguez:2024jzw,Long:2024tws}.

In order to analyze the energy extraction, it is vital to know the outflow speed of plasmoids. Most of previous works adopted the description of Sweet-Parker model \cite{SP1,SP2,Lyubarsky2006}, the first analytical model under the scheme of magnetohydrodynamics (MHD), which showed that the outflow speed just equals the Alfv{\'e}n velocity. However, the extremely low reconnection rates predicted by the Sweet-Parker model contrast sharply with the observational results found in most astrophysical system \cite{yamada2009}. Another model was proposed by Petchek to increase the reconnection rate, by setting an open outflow region \cite{Petschek}. Although its status as the first analytical model of fast reconnection is significant, its description is complicated and its self-consistency was challenged \cite{Biskamp1986,Sato1979}. In Ref.~\cite{Liu2017}, another analytical model of magnetic reconnection was proposed under the MHD scheme. It predicts that the maximal local reconnection rate is of order 0.1, which is consistent with the observations \cite{2002ApJ...565.1335Q,Isobe:2005ub}. Compared to Sweet-Parker model, model in Ref.~\cite{Liu2017} only induced one additional parameter: the geometric index, defined to be the ratio of width to length of the current sheet. Moreover, when the geometric index approaches zero, this model reduces to the Sweet-Parker configuration spontaneously. 

Alongside analytical models, numerical simulation is another robust approach to study the magnetic reconnection \cite{Spitkovsky:2005vsy,2011PhRvL107b5002K,2015PhRvL115q5004S,Jia:2023iup}. Numerical simulations can effectively account for complex and chaotic systems, which are common in astrophysical scenarios. Using the particle-in-cell (PIC) method \cite{PIC1,PIC2,PIC3,PIC4,PIC5}, simulations are initialized by setting anti-parallel magnetic field lines on either side of a thin current sheet. Unlike most analytical models, the processes of magnetic reconnection described numerically were not limited to a two-dimensional plane. A three-dimensional process would be vastly different from the two-dimensional models in terms of plasma kinetics and thermodynamics \cite{Comisso:2023ygd}. 

In Ref.~\cite{CA2021}, the analytical model from Ref.~\cite{Liu2017} was adopted to analyze energy extraction via magnetic reconnection. However, as they set an infinitesimal geometric index, results were actually equivalent to those obtained by adopting the Sweet-Parker model. In Ref.~\cite{Work0} and Ref.~\cite{Work1}, the authors adopted the analytical model introduced in Ref.~\cite{Liu2017} as well but opted for a geometric index that maximizes the local reconnection rate in the limit of high magnetization. However, the value of geometric index should, in principle, vary randomly, even if the magnetic reconnection occurs as what the analytical model in Ref.~\cite{Liu2017} predicted in nature.

In this work, we aim to extend the analyses in Ref.~\cite{Work1} by derestricting the parameters in reconnection models, such as the geometric index in the analytical model introduced in Ref.~\cite{Liu2017} and the guide field fraction in a three-dimensional model coming from numerical works. Align with the discussions in Ref.~\cite{Work1}, the energy extraction efficiencies and the allowed regions for energy extraction in parameter planes are exhibited. Additionally, we calculate the covering factor introduced in Ref.~\cite{Work1}, which quantifies the capability of an accretion system in extracting energy from a rotating black hole via magnetic reconnection. We argue that it is unsuitable to simply select certain values for the parameters in reconnection models when considering energy extraction, as different choices of parameters lead to different consequences. As progress, we present the distributions of covering factor as functions of both black hole spin and either the geometric index or guide field fraction, from which one could easily determine the minimally allowed values of parameters for a given black hole spin or conversely the minimally allowed black hole spin after fixing the parameters. Moreover, we recommend to define the averaged covering factor weighted by reconnection rate, which helps us quantify the capability of an accretion system in energy extraction via magnetic reconnection in an overall manner under the assumption that the magnetic reconnection occurs stochastically and randomly in astrophysical systems.

The remain parts of the paper is organized as follows. In Sect.~\ref{sec:concept}, we briefly recap some basic concepts about energy extraction via Comisso-Asenjo process. The two reconnection models we employ in this work are introduced in Sect.~\ref{sec:Liu} and Sect.~\ref{sec:3D}. We present our results in Sect.~\ref{sec:result}, including the energy extraction efficiencies and the allowed regions for energy extraction in Sect.~\ref{sec:para}, the covering facor in Sect.~\ref{sec:chi} and the averaged covering factor weighted by reconnection rate in Sect.~\ref{sec:ave-chi}. In Sect.~\ref{sec:sum}, we summarize this work. In the following discussions, we set $G=M=c=1$ without losing generality.

\section{Basic concepts}
\label{sec:concept}

In this section, let us revisit some basic concepts introduced in Ref.~\cite{Work1} for quantifying the energy extraction via magnetic reconnection, via Comisso-Asenjo process specifically. Magnetic reconnection is considered to occur in a stationary, axisymmetric spacetime, in which the line element could be represented in 3+1 formalism as \cite{MacDonald:1982zz}:
\begin{equation}
    ds^2=g_{\mu\nu}dx^{\mu}dx^{\nu}=-\alpha^2dt^2+\sum_{i=1}^3\left(h_idx^i-\alpha\beta^idt\right)^2
    \label{eq:line}
\end{equation}
where $\alpha$ is the lapse function, $h_i$ is the scale factor and $\beta^i=h_i\omega^i/\alpha$ is the shift vector with $\omega^i$ being the velocity of frame dragging. We focus on energy extraction from a Kerr black hole, such that:
\begin{equation}
    \alpha=\sqrt{\frac{\Delta\Sigma}{A}},~~h_r=\sqrt{\frac{\Sigma}{\Delta}},~~h_{\theta}=\sqrt{\Sigma},~~h_{\phi}=\sqrt{\frac{A}{\Sigma}}\sin\theta,~~
    \omega^r=\omega^{\theta}=0,~~\omega^{\phi}=\frac{2ar}{A}
    \label{eq:alpha_h_omega}
\end{equation}
in BL coordinates $\left(t,r,\theta,\phi\right)$, with $\Sigma=r^2+a^2\cos^2\theta$, $\Delta=r^2-2r+a^2$ and $A=\left(r^2+a^2\right)^2-a^2\Delta\sin^2\theta$. 

Magnetic reconnection is assumed to happen in the bulk plasma which is treated as ideal fluid flowing on the equatorial plane. It is necessary to quantify the process in the fluid's rest frame and that could be defined via the normal tetrad as 
\begin{equation}
    \begin{split}
        e_{[0]}^{\mu}&=\hat{\gamma}_s \left[\hat{e}_{(t)}^{\mu}+\hat{v}_s^{(r)}\hat{e}_{(r)}^{\mu}+\hat{v}_s^{(\phi)}\hat{e}_{(\phi)}^{\mu}\right], \\
        e_{[1]}^{\mu}&=\frac{1}{\hat{v}_s}\left[\hat{v}_s^{(\phi)}\hat{e}_{(r)}^{\mu}-\hat{v}_s^{(r)}\hat{e}_{(\phi)}^{\mu}\right],~~
        e_{[2]}^{\mu}=\hat{e}_{(\theta)}^{\mu}, \\
        e_{[3]}^{\mu}&=\hat{\gamma}_s\left[\hat{v}_s\hat{e}_{(t)}^{\mu}+\frac{\hat{v}_s^{(r)}}{\hat{v}_s}\hat{e}_{(r)}^{\mu}+\frac{\hat{v}_s^{(\phi)}}{\hat{v}_s}\hat{e}_{(\phi)}^{\mu}\right],
    \end{split}
    \label{eq:fluid}
\end{equation}
where $\hat{v}_s=\sqrt{\left(\hat{v}_s^{(r)}\right)^2+\left(\hat{v}_s^{(\phi)}\right)^2}$ is the speed of fluid with $\hat{v}_s^{(r)}$ and $\hat{v}_s^{(\phi)}$ being the components of 3-velocity and $\hat{\gamma}_s$ is the Lorentz factor, observed by zero-angular-momentum observers (ZAMOs). While $\hat{e}^{\mu}_{(\nu)}$ is the normal tetrad of ZAMOs. One can see that $e_{[1]}$ and $e_{[3]}$ are orthogonal and parallel to the moving direction of fluid, respectively.

As shown in numerical simulations, the rotation of black hole may produce antiparallel magnetic field lines adjacent to the equatorial plane \cite{Ripperda2020,Yuan2024-2}, which makes the existence of an equatorial current sheet possible. The equilibirium of anti-parallel magnetic field would be broken by plasmoid instability \cite{Comisso:2016pyg,Comisso:2017arh,PhysRevLett.121.165101}, which subsequently actuates fast magnetic reconnection \cite{Daughton2009} and then expels created plasmoids away. Upon the plasmoid instability occurring near the black hole, the Comisso-Asenjo process assumed two plasmoids ejected oppositely in the local rest frame of bulk plasma \cite{CA2021}. The 4-velocities of ejected plasmoids could be writtern as \cite{Work0}:
\begin{equation}
    u^{\mu}_{\pm}=\gamma_{\rm out}\left[e_{[0],0}^{\mu}\pm 
    v_{\rm out}\left(\cos\xi_{B}e_{[3],0}^{\mu}+\sin\xi_{B}e_{[1],0}^{\mu}\right)\right]
    \label{eq:u_out}
\end{equation}
where $v_{\rm out}$ is the outflow speed observed in the fluid's rest frame while $\gamma_{\rm out}$ is the Lorentz factor. The subscript ",0" represents the reconnection point and "$\pm$" represents two plasmoids ejected toward opposite directions viewed in the fluid's rest frame. The orientation angle is represented by $\xi_B$, regarded as a free parameter ranging from $-\pi/2$ to $\pi/2$. The outflow speed generally depends on the local magnetization $\sigma_0$ on the reconnection point, whose explicit expression corresponds to the local configuration of magnetic field and how the magnetofluid diffuses from the current sheet. Analytical forms of $v_{\rm out}$ would be different if different models are applied. We will introduce the reconnection models in Sect.~\ref{sec:model} detailedly and demonstrate how $v_{\rm out}$ depends on the local magnetization and parameters in different models.

Efficiency of energy extraction is defined to be \cite{CA2021}
\begin{equation}
    \eta=\frac{\epsilon_+}{\epsilon_++\epsilon_-}
    \label{eq:eta}
\end{equation}
where $\epsilon_{\pm}$ are the energy-at-infinity per enthalpy of ejected plasmoids, satisfying \cite{CA2021,Work0}
\begin{equation}
    \begin{split}
        \epsilon_{\pm}=&-u_t^{\pm}-\frac{\Tilde{p}_{\pm}}{u^t_{\pm}} \\
        =&\alpha\hat{\gamma}_s\gamma_{\rm out}\left[\left(1+\beta^{\phi}\hat{v}_s^{(\phi)}\right)\pm 
        v_{\rm out}\left(\hat{v}_s+\beta^{\phi}\frac{\hat{v}_s^{(\phi)}}{\hat{v}_s}\right)\cos\xi_B\mp
        v_{\rm out}\beta^{\phi}\frac{\hat{v}_s^{(r)}}{\hat{\gamma}_s\hat{v}_s}\sin\xi_B\right] \\
        &-\frac{\alpha\Tilde{p}}{\hat{\gamma}_s\gamma_{\rm out}\left(1\pm \hat{v}_s v_{\rm out}\cos\xi_B\right)}.
    \end{split}
    \label{eq:epsilon}
\end{equation}
We adopt $\Tilde{p}_{\pm}=1/4$ for treating the plasmoids to be ultra-relativistic \cite{Lyubarsky2006}.

Since we assume that the bulk plasma flows along timelike geodesics on the equatorial plane, its 4-velocity $U^{\mu}$, when projected onto the normal tetrad of ZAMOs, reads:
\begin{equation}
    U^{\mu}\hat{e}^{(a)}_{\mu}=\hat{\gamma}_s\left\{1,\hat{v}_s^{(r)},0,\hat{v}_s^{(\phi)}\right\}=
    \left\{\frac{E-\omega^{\phi}L}{\alpha},h_rU^r,0,\frac{L}{h_{\phi}}\right\}
    \label{eq:U}
\end{equation}
where $E$ and $L$ are the two conserved quantities. Two kinds of timelike geodesics are under consideration. The first one is the circular orbits (prograde only), on which $U^r=0$ and
\begin{equation}
    E=E_{\rm K}(r)=\frac{r^{3/2}-2r^{1/2}+a}{\sqrt{r^3-3r^2+2ar^{3/2}}},~~
    L=L_{\rm K}(r)=\frac{r^2-2ar^{1/2}+a^2}{\sqrt{r^3-3r^2+2ar^{3/2}}}
    \label{eq:E_L}
\end{equation}
This orbit exists from the infinity down to the photon sphere $r_{\rm ph}$. Another one is the plunging geodesic (also prograde only), on which \cite{Mummery2022}
\begin{equation}
    E_{\rm ms}=E_{\rm K}\left(r_{\rm ms}\right),~~L_{\rm ms}=L_{\rm K}\left(r_{\rm ms}\right)
    \label{eq:E_L_plunge}
\end{equation}
and
\begin{equation}
    U^r=-\sqrt{\frac{2}{3r_{\rm ms}}}\left(\frac{r_{\rm ms}}{r}-1\right)^{3/2}
    \label{eq:Ur_plunge}
\end{equation}
This orbit exists from the innermost stable circular orbit (ISCO) $r_{\rm ms}$ down to the event horizon $r_{\rm EH}$. Align with the discussions in Ref.~\cite{Work1}, we analyze the magnetic reconnection occuring in bulk plasma with two kinds of streamlines. The first is refered to as the combined streamline, in which the bulk plasma flows circlular outside ISCO before plunging into the black hole. The second is refered to as the circular streamline, where the bulk plasma flows circularly all the way down to the photon sphere. After obtaining the information of $E$, $L$ and $U^r$, one can get the 3-velocity of bulk plasma viewed by ZAMOs, which are essential for calculating the energy-at-infinity per enthalpy and the efficiency of energy extraction based on Eq.~\ref{eq:epsilon} and \ref{eq:eta} respectively.

\section{Reconnection models}
\label{sec:model}

In this section we revisit two reconnection models we employ in this work. The first one is an analytical model under MHD scheme which was proposed in Ref.~\cite{Liu2017}. The second one came from numerical simulations by PIC method.

\subsection{Anti-parallel reconnection model: geometric index}
\label{sec:Liu}

\begin{figure}
    \centering
    \includegraphics[width=\textwidth]{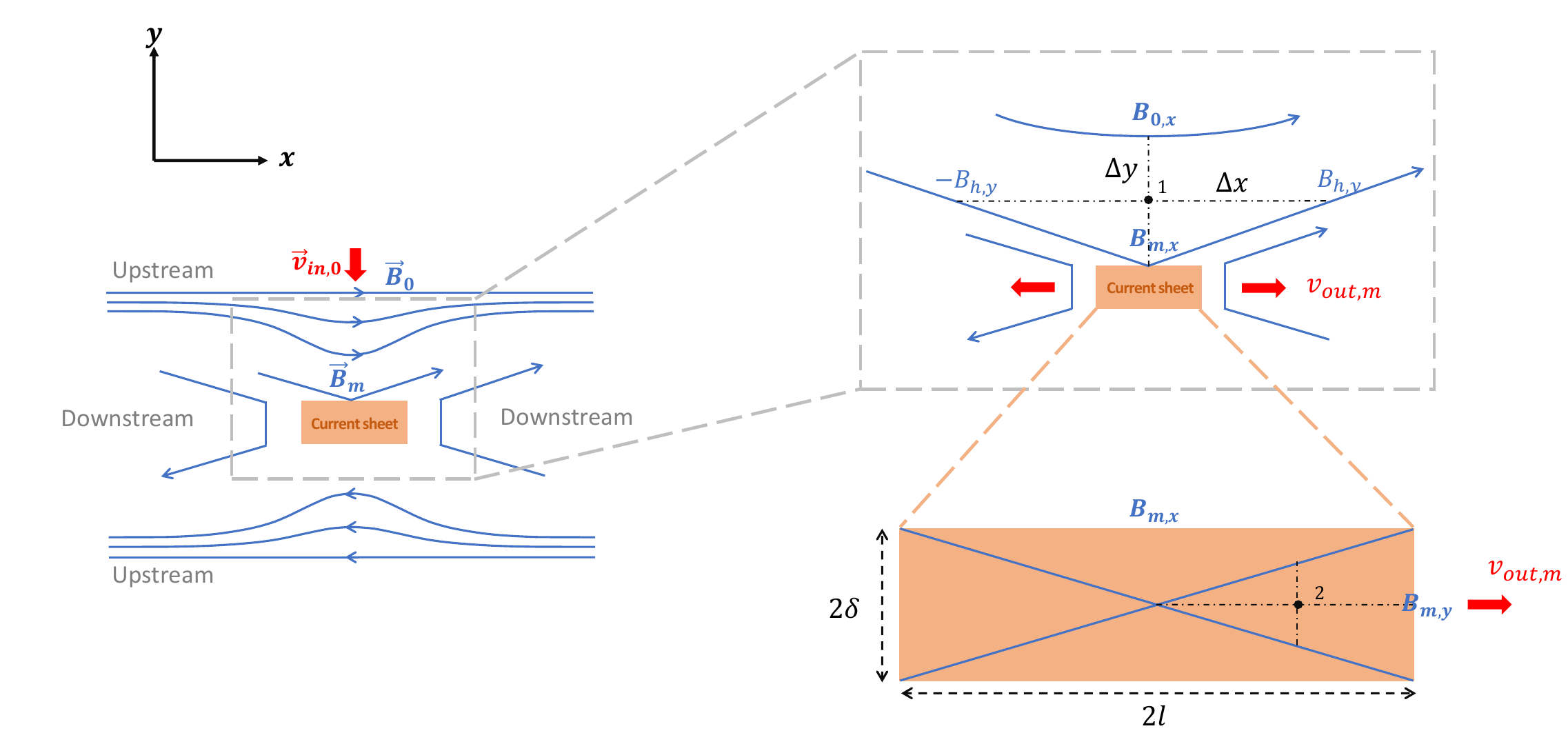}
    \caption{Schematic diagram of the anti-parallel model in local scale (left and top right panels) and microscale (bottom right panel). The blue lines represent the magnetic field while the red arrows represent the flowing direction of magnetofluid. The orange square represents the current sheet.}
    \label{fig:Liu}
\end{figure}

%Liu's model was proposed as a fast, quasi-stationary reconnection model, designed to explain the local reconnection rate in solar flares and substorm in magnetotails of Earth, under the scheme of MHD. We choose the name of the first author of Ref.~\cite{Liu2017} to address this model temporarily.

Now let us revisit the analytical model of magnetic reconnection introduced in Ref.~\cite{Liu2017} in detail. As a two dimensional model, it provides a concise description for the fast reconnection analytically and was frequently applied in the researches about Comisso-Asenjo process. We henceforth referred this model to be the anti-parallel model, in opposed to the three dimensional model introduced in Sect.~\ref{sec:3D}. 

The schematic diagram is plotted in Fig.~\ref{fig:Liu}, where the blue lines represent the magnetic field lines. The left and top right panels show the structure of magnetic field at the local scale, where the mass density, pressure and enthalpy of the magnetofluid reach uniformity. What we generally call the local magnetization is just the magnetization upstream at the local scale, characterized by two branches of anti-parallel magnetic field lines approaching each other. The orange square in Fig.~\ref{fig:Liu}, or the bottom right panel, represent the current sheet (or diffusion region), within which the anti-parallel magnetic lines touch and reconnect. We use the subscript "0" to denote physical quantities upstream while "$m$" to denote those on the surfaces of the current sheet. The geometric index is defined to be \cite{Work0}:
\begin{equation}
    {\sf g}=\frac{\Delta y}{\Delta x}=\frac{\delta}{l}
    \label{eq:geo}
\end{equation}
where $\Delta y$ and $\Delta x$ are the half thickness and half length of the local scale, while $\delta$ and $l$ are the half thickness and half length of the current sheet, respectively. The configuration of the magnetic field at the local scale is assumed to satisfy:
\begin{equation}
    \frac{B_{m,y}}{B_{m,x}}=\frac{2B_{h,y}}{B_{0,x}+B_{m,x}}={\sf g}
    \label{eq:geo1}
\end{equation}
where "$h$" denotes the points on the two sides of the point 1, marked on the top right panel in Fig.~\ref{fig:Liu}.

On point 1, if the magnetic field retains its structure, the gradient of magnetic pressure, $\nabla B^2/2$, should be balanced by the magnetic tension, $(\Vec{B}\cdot\nabla)\Vec{B}$. This indicates:
\begin{equation}
    \frac{B_{0,x}^2-B_{m,x}^2}{2\Delta y}\simeq \left(\frac{B_{0,x}+B_{m,x}}{2}\right)\times\frac{1}{\Delta x}\times\left[B_{h,y}-(-B_{h,y})\right]
    \label{eq:eq1}
\end{equation}
where the $x$-component of magnetic field on point 1 is assumed to be $\left(B_{0,x}+B_{m,x}\right)/2$. Applying the relations in Eq.~\ref{eq:geo1}, one gets:
\begin{equation}
    \frac{B_{m,x}}{B_{0,x}}=\frac{1-{\sf g}^2}{1+{\sf g}^2}
    \label{eq:Bm0}
\end{equation}
Similarly, the balance on the point 2, marked on the bottom right panel in Fig.~\ref{fig:Liu}, yields:
\begin{equation}
    \frac{\omega\gamma_{{\rm out},m}^2v_{{\rm out},m}^2}{2l}+\frac{B_{m,y}^2}{2l}\simeq
    \frac{B_{m,y}}{2}\times\frac{1}{\delta}\times\frac{B_{m,x}-(-B_{m,x})}{2}
    \label{eq:point2}
\end{equation}
where $\omega$ is the local enthalpy of plasma. 

Assuming that plasmoids are not decelerated after being ejected from the current sheet, one can obtain the following relation from Eq.~\eqref{eq:point2}:
\begin{equation} 
    v_{\rm out}\simeq v_{{\rm out},m}\simeq \sqrt{\frac{\left(1-{\sf g}^2\right)\sigma_{m,x}}{1+\left(1-{\sf g}^2\right)\sigma_{m,x}}} \simeq \sqrt{\frac{\left(1-{\sf g}^2\right)^3\sigma_0}{\left(1+{\sf g}^2\right)^2+\left(1-{\sf g}^2\right)^3\sigma_0}} 
    \label{eq:voutL} 
\end{equation}
where $\sigma_m$ and $v_{{\rm out},m}$ are the magnetization and outflow speed on the sides of the current sheet respectively, while $\sigma_0=B_{0,x}^2/\omega$ is the magnetization upstream (identical to the local magnetization). It can be easily seen from Eq.~\eqref{eq:voutL} that the anti-parallel model reduces to the Sweet-Parker configuration \cite{SP1,SP2,Lyubarsky2006} whenever ${\sf g}\rightarrow 0$  (hence $\delta\ll l$, consistent with the assumptions of the Sweet-Parker model). Moreover, based on  Eq.~\eqref{eq:voutL}, the condition ${\sf g}<1$ (thus $\delta<L$) is required for magnetic reconnection to occur.

The reconnection rate in this model satistfies \cite{Work0}:
\begin{equation}
    R_0\equiv \frac{v_{\rm in,0}}{v_{{\rm A},0}}\simeq \frac{B_{m,y}}{B_{m,x}}\times\frac{B_{m,x}}{B_{0,x}}\times\frac{v_{{\rm out},m}}{v_{{\rm A},0}}
    \simeq {\sf g}\frac{1-{\sf g}^2}{1+{\sf g}^2}
    \sqrt{\frac{\left(1-{\sf g}^2\right)^3\sigma_0}{\left(1+{\sf g}^2\right)^2+\left(1-{\sf g}^2\right)^3\sigma_0}}
    \label{eq:R0L}
\end{equation} 
where the Ohm's law
\begin{equation}
    E_z\simeq B_{m,y}v_{\rm out}\simeq B_{0,x}v_{\rm in,0}
\end{equation}
is applied. Here, $v_{{\rm A},0}$ is the Alfv{\'e}n velocity upstream (also referred to as the local Alfv{\'e}n velocity). We adopt the high magnetization limit ($\sigma_0\gg 1$) to simplify the expression. Similar to the Petschek model \cite{Petschek}, the anti-parallel model never requires the magnetofluid coming upstream to be entirely ejected from two sides of current sheet. As a result, it is capable of describing the magnetic reconnection in a fast local reconnection rate. One can see from Eq.~\eqref{eq:R0L} that the reconnection rate is maximized at ${\sf g}=0.49$, with the maximal value reaching about 0.3, in high magnetization limit \cite{Work0}. While this model is more analyzable than the Petschek model because—compared to the Sweet-Parker model—only one additional parameter, the geometric index, is induced to fully describe the magnetic reconnection process. We will see in Sect.~\ref{sec:para} that the increase of geometric index slows down the outflow speed of plasmoids, which ultimately increases the difficulty of energy extraction via magnetic reconnection within the ergosphere of a rotating black hole.

\subsection{Quasi-3D reconnection model: guide field}
\label{sec:3D}

%\begin{figure}
%    \centering
%    \includegraphics[width=0.83\textwidth]{picg.pdf}
%    \caption{Schematic diagram of Harris equilibrium with guide field, viewed in $x-y$ (left panel) and $x-z$ (right panel) planes. The blue lines represent the magnetic field, while the orange square represents the current sheet. The dashed lines in the right panel means that the magnetic field lines are located beneath the current sheet.}
%    \label{fig:Guide}
%\end{figure}
%\begin{figure}
%    \centering
%    \includegraphics[width=\textwidth]{3D.pdf}
%    \caption{Schematic diagram of magnetic reconnection when guide field is added initially. The magnetic field lines are represented by blue or green lines. Lines with same colors are connected. The left panel shows the magnetic structure in the local scale, with orange square representing the microscale. The right panel depicts the evolution of magnetic field at microscale, where two branches of magnetic field lins approach, reconnect at point $X$ and eject plasmoids.}
%    \label{fig:3D}
%\end{figure}

Most analytical models of magnetic reconnection are two-dimensional, in which the anti-parallel magnetic field lines are confined to $x-y$ plane. There were some studies trying to extend the magnetic field to the third dimension when analyzing the reconnection \cite{Lyubarsky2006} but no surprising or analyzable result have been obtained. Conversely, numerical simulations have yielded more promising results when non-zero $z$-component of magnetic field was added. Generally, in PIC simulations, the magnetic field is initialized based on the Harris sheet equilibrium \cite{Harris1962}, which established a steady state formed by two branches of anti-parallel magnetic field lines on two sides of the current sheet in $x-y$ plane. After adding $z$-component of magnetic field, the initial magnetic structure reads
\begin{equation}
    \Vec{B}=B_{\parallel}\tanh(y/\delta)\hat{x}+B_g\hat{z}
    \label{eq:iniB}
\end{equation}
where $B_g$ is usually called the guide field. 

Based on PIC simulations, reconnection occurs when tiny perturbations are applied to the Harris sheet equilibrium with the guide field. We defined the guide field fraction as:
\begin{equation}
    \mathcal{G}=\frac{B_g^2}{B_{\parallel}^2+B_g^2}=\frac{\sigma_g}{\sigma_{\parallel}+\sigma_g}=\frac{\sigma_g}{\sigma_0}
    \label{eq:fraction}
\end{equation}
where $\sigma_{\parallel}=B_{\parallel}^2/\omega$ and $\sigma_g=B_g^2/\omega$. Both $B_{\parallel}$ and $B_g$ should be considered when calculating the local magnetization $\sigma_0$. Based on the simulations, the outflow speed of plasmoids, which are ejected along $x$-axis as well, should satisfy \cite{Comisso:2023ygd}:
\begin{equation}
    v_{\rm out}\simeq \sqrt{\frac{\sigma_{\parallel}}{1+\sigma_{\parallel}+\sigma_g}}=v_{\rm A,0}\sqrt{1-\mathcal{G}}
    \label{eq:voutg}
\end{equation}
where $v_{\rm A,0}=\sqrt{\sigma_0/(1+\sigma_0)}$ is the local Alfv{\'e}n velocity. The presence of the guide field decreases the outflow speed, which makes the conditions for energy extraction within the ergosphere of a rotating black hole more restrictive. The local reconnection rate would approximately obey: \cite{Comisso:2023ygd}:
\begin{equation}
    R_0\simeq \Tilde{R}_0\sqrt{\frac{1+\sigma_{\parallel}}{1+\sigma_{\parallel}+\sigma_g}}
    =\Tilde{R}_0\sqrt{\frac{\mathcal{G}+\sigma_0(1-\mathcal{G})}{\mathcal{G}+\sigma_0}}
    \label{eq:R0g}
\end{equation}
Here $\Tilde{R}_0$ represents the magnitude of the local reconnection rate in the absence of the guide field, which is usually chosen to be 0.1. The reconnection rate also decreases with the increase of guide field fraction. In the following discussion, the reconnection model described in this section will be referred to as the quasi-3D model. In Sect.~\ref{sec:para}, we will examine how the energy extraction via magnetic reconnection depends on the guide field fraction in detail.

\section{Results}
\label{sec:result}

In this section, we present our results. We show the efficiencies of energy extraction via magnetic reconnection for multiple values of geometric index or guide field fraction, respectively. Additionally, the allowed regions for energy extraction (regions with $\eta>1$) in parameter planes consisting of reconnection point and orientation angle are plotted. Furthermore, we exhibit the distribution of covering factor for energy extraction, introduced in Ref.~\cite{Work1}, in the parameter planes of black hole spin and either geodesic index or guide field fraction. Finally, we recommend to calculate the averaged covering factor weighted by reconnection rate.

\subsection{Dependences on parameters}
\label{sec:para}

\begin{figure}
    \centering
    \hspace{-2.5mm}
    \includegraphics[width=0.48\textwidth]{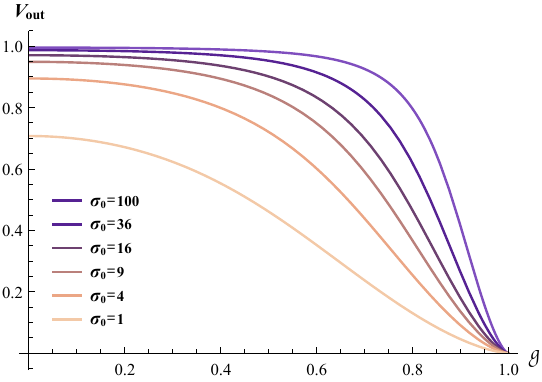}
    \hspace{2.5mm}
    \includegraphics[width=0.48\textwidth]{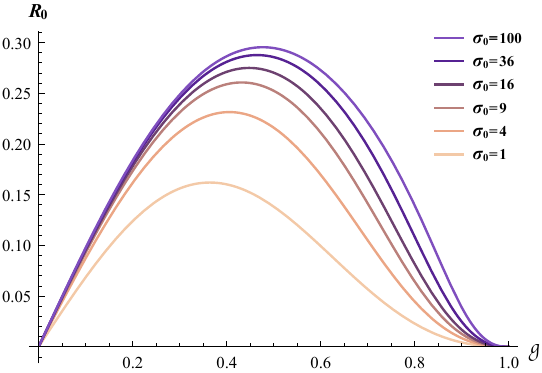}
    \hspace{-2.5mm}
    \caption{Outflow speed (left panel) and reconnection rate (right panel) as functions of the geometric index in anti-parallel model.}
    \label{fig:vout-R0-L}
\end{figure}

We plot the outflow speed $v_{\rm out}$ and local reconnection rate $R_0$\footnote{In Fig.~\ref{fig:vout-R0-L}, we still adopt Eq.~\eqref{eq:R0L} for simplicity when the local magnetization is not too large.} in the anti-parallel model as functions of geometric index for multiple values of local magnetization in Fig.~\ref{fig:vout-R0-L}. We can see from the left panel that the outflow speed decreases with the increase of geometric index. The variation is smooth when the local magnetization is low. However, in high magnetization limit, it approaches a step function. In contrast to $v_{\rm out}$, the local reconnection rate reaches its maximum when ${\sf g}\simeq 0.3\sim0.5$. The extreme point of reconnection rate locates at ${\sf g}\simeq 0.49$ when $\sigma_0\gg 1$, with maximal value of $R_0$ to be around 0.3.

\begin{figure}
    \centering
    \includegraphics[width=0.45\textwidth]{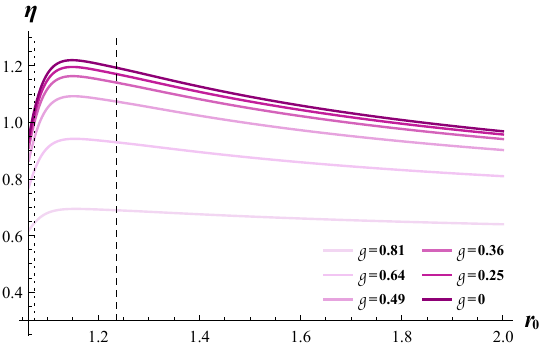}
    \hspace{2.5mm}
    \includegraphics[width=0.45\textwidth]{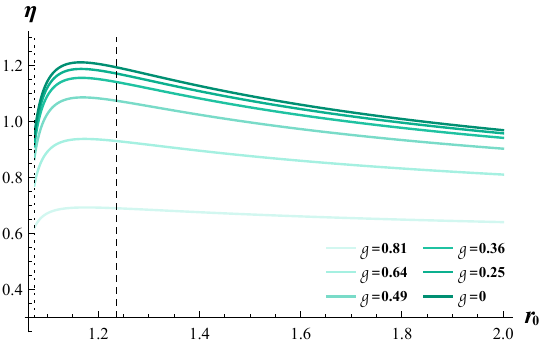}
    \hspace{-2.5mm}
    \includegraphics[width=0.45\textwidth]{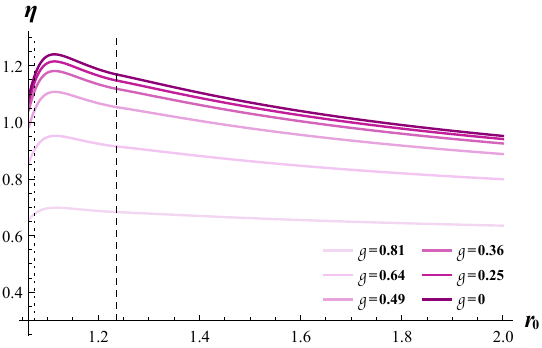}
    \hspace{2.5mm}
    \includegraphics[width=0.45\textwidth]{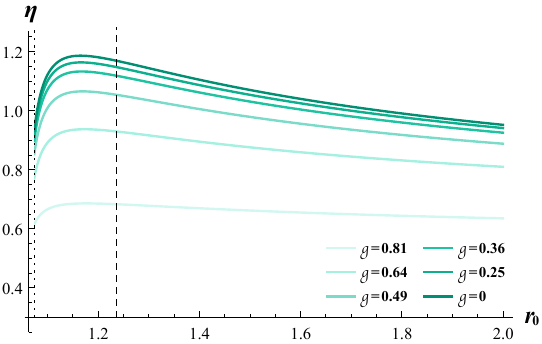}
    \hspace{-2.5mm}
    \includegraphics[width=0.45\textwidth]{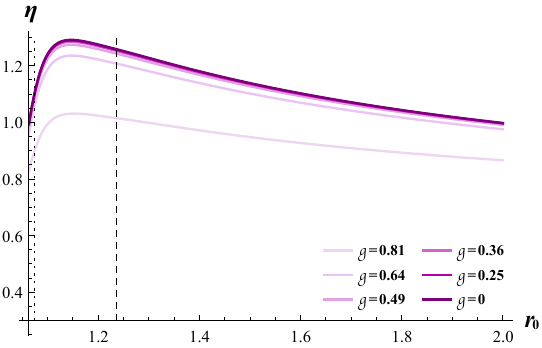}
    \hspace{2.5mm}
    \includegraphics[width=0.45\textwidth]{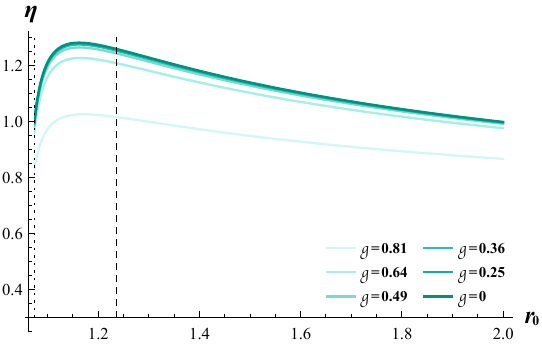}
    \hspace{-2.5mm}
    \includegraphics[width=0.45\textwidth]{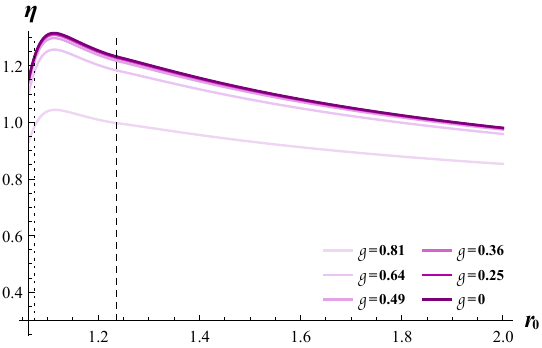}
    \hspace{2.5mm}
    \includegraphics[width=0.45\textwidth]{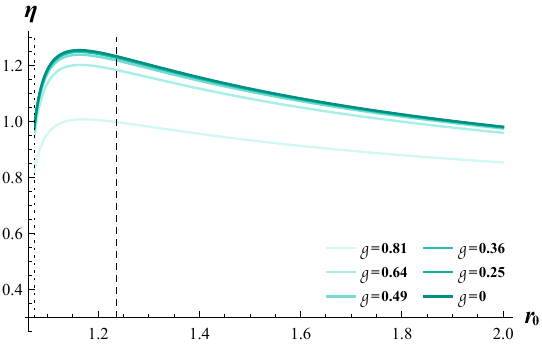}
    \caption{Efficiencies of energy extraction for multiple values of geometric index in the anti-parallel model with $a=0.998$, $\sigma_0=4$ (panels on first and second rows) and $\sigma_0=50$ (panels on third and fourth rows), $\xi_B=0$ (panels on first and third rows) and $\xi_B=\pi/12$ (panels on second and fourth rows), combined streamline (left panels) and circular streamline (right panels). The dotted and dashed black lines represent the photon sphere and ISCO, respectively.}
    \label{fig:eta-L}
\end{figure}

The efficiencies of energy extraction as functions of radius of the reconnection point $r_0$ for multiple values of geometric index are plotted in Fig.~\ref{fig:eta-L}, with $a=0.998$, $\xi_B=0$ and $\xi_B=\pi/12$, $\sigma_0=4$ and $\sigma_0=50$, combined and circular streamlines. The radius of ISCO and that of the photon sphere are (henceforth) represented by dashed and dotted black lines, respectively. We can see that the efficiencies of energy extraction via magnetic reconnection for the two types of streamlines are similar when $\xi_B=0$. However, in the plunging region ($r<r_{\rm ms}$), energy extraction becomes much more efficient in the plunging bulk plasma when $\xi_B=\pi/12>0$, as shown in the second and fourth rows of panels in Fig.~\ref{fig:eta-L}. It is consistent with the results in Ref.~\cite{Work1}. Energy extraction is less efficient when ${\sf g}$ is higher, as a larger geometric index results in a lower outflow speed. While in the limit of high magnetization, as mentioned before, the outflow speed behaves like a step function with respect to the geometric index. Hence the influence of geometric index is not significant unless ${\sf g}$ approaches 1, as shown in the bottom two rows of panels in Fig.~\ref{fig:eta-L}.

\begin{figure}
    \centering
    \includegraphics[width=0.48\textwidth]{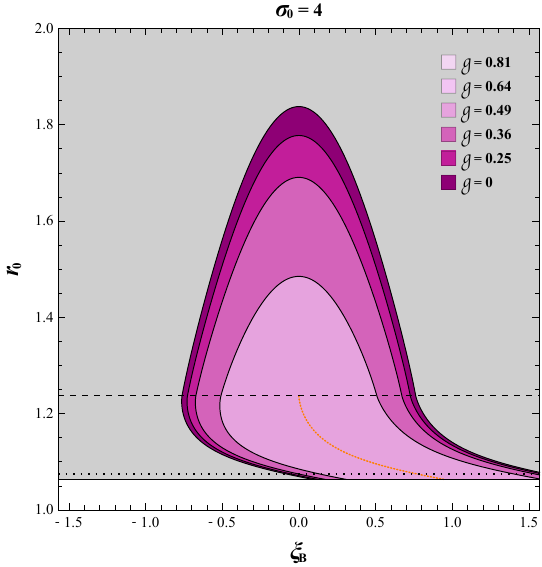}
    \hspace{3mm}
    \includegraphics[width=0.48\textwidth]{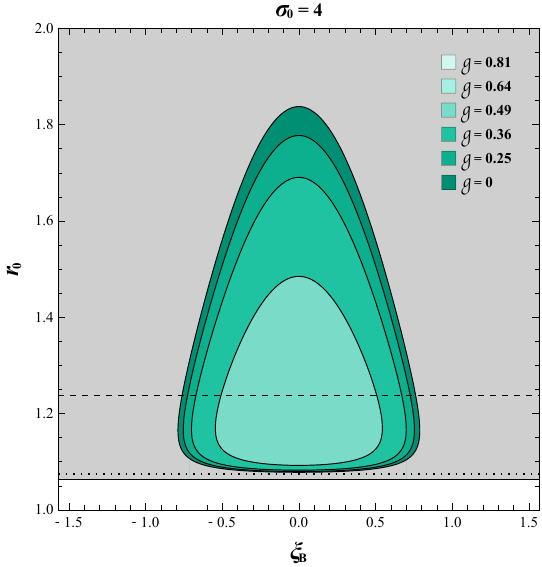}
    \hspace{-3mm}
    \includegraphics[width=0.48\textwidth]{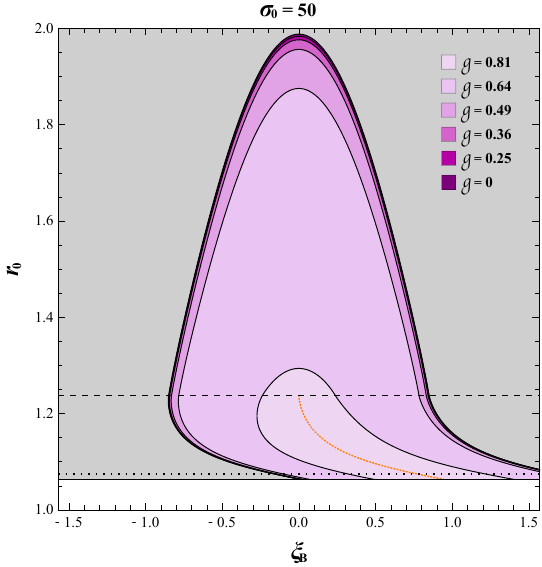}
    \hspace{3mm}
    \includegraphics[width=0.48\textwidth]{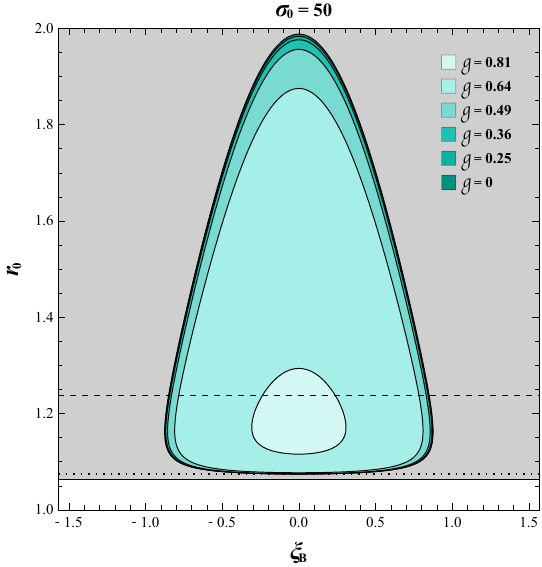}
    \caption{The allowed regions for energy extraction in $\xi_B-r_0$ planes with combined streamline are colored in plum (left panels) and those with circular streamline are colored in turquoise (right panels) for multiple values of geometric index in the anti-parallel model,  with $a=0.998$, $\sigma_0=4$ (upper panels) and $\sigma_0=50$ (lower panels). The orange dotted lines in plunging regions of the left panels represent the best orientation angles in the limit of high local magnetization for plunging bulk plasma. The solid, dotted and dashed black lines represent event horizon, photon sphere and ISCO respectively.}
    \label{fig:r0-xi-L}
\end{figure}

We plot the allowed regions for energy extraction in $\xi_B-r_0$ planes in Fig.~\ref{fig:r0-xi-L} for multiple values of geometric index, with $a=0.998$, $\sigma_0=4$ and $\sigma_0=50$, combined and circular streamlines. The event horizon is (henceforth) represented by solid straight lines. We can see that for circularly flowing bulk plasma, the allowed regions are just symmetric about $\xi_B=0$. While for plunging bulk plasma, the allowed regions incline to the regions with positive orientation angle. We plot the best orientation angle in the limit of high local magnetization \cite{Work1},
\begin{equation}
        \xi_B=\xi_{B,{\rm m}}\equiv \arctan\left(-\frac{\beta^{\phi}\hat{v}_s^{(r)}}{\hat{\gamma}_s\hat{v}_s^2+\beta^{\phi}\hat{\gamma}_s\hat{v}_s^{(\phi)}}\right)
    \label{eq:xi_m}
\end{equation}
which is defined to be the orientation angle that maximizes the energy extraction efficiency, by orange dotted lines in the plunging region on left panels. We observe that the inclinations of the allowed regions, when the bulk plasma plunges, tend to follow the lines of the best orientation angle. This is consistent with the results in Ref.~\cite{Work1}.

\begin{figure}
    \centering
    \includegraphics[width=0.48\textwidth]{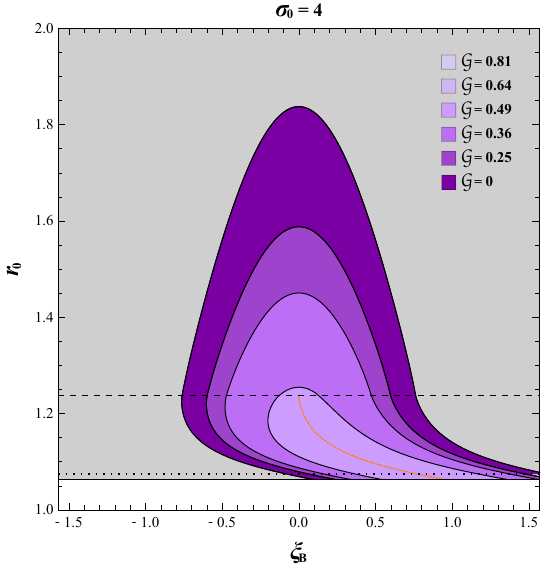}
    \hspace{3mm}
    \includegraphics[width=0.48\textwidth]{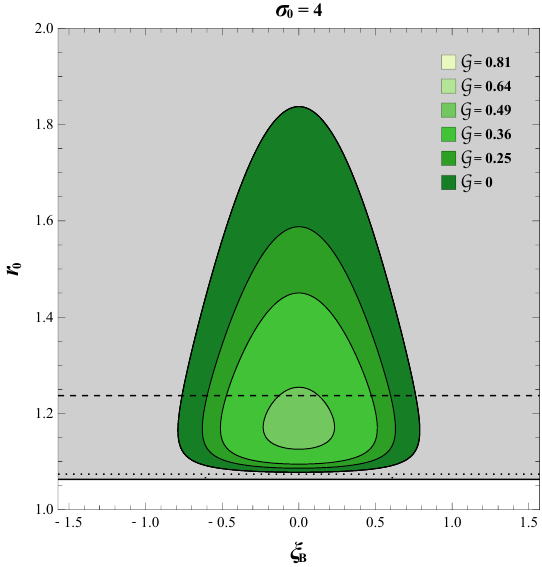}
    \hspace{-3mm}
    \includegraphics[width=0.48\textwidth]{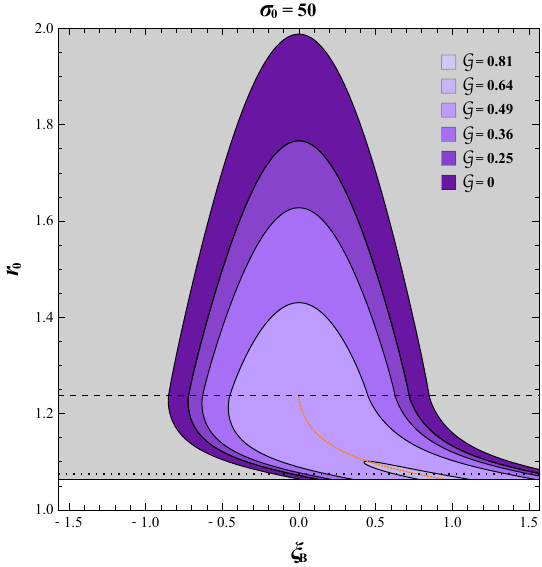}
    \hspace{3mm}
    \includegraphics[width=0.48\textwidth]{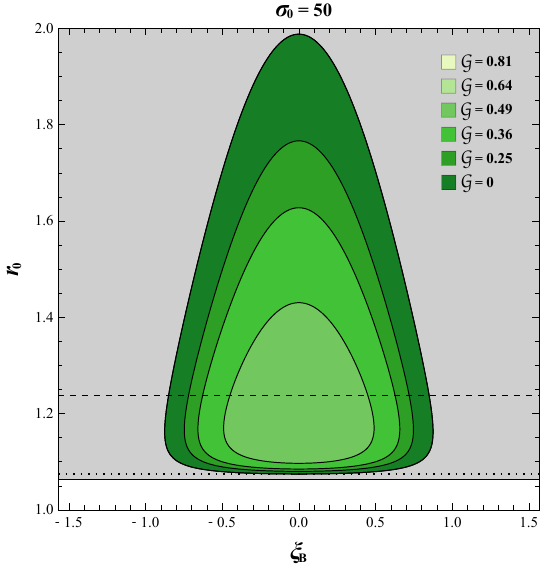}
    \caption{The allowed regions for energy extraction in $\xi_B-r_0$ planes with combined streamline are colored in violet (left panels) and those with circular streamline are colored in green (right panels) for multiple values of geometric index in  quasi-3D model, with $a=0.998$, $\sigma_0=4$ (upper panels) and $\sigma_0=50$ (lower panels). The orange dotted lines in plunging region of the left panels represent best orientation angles in the limit of high local magnetization for plunging bulk plasma. The solid, dotted and dashed black lines represent event horizon, photon sphere and ISCO, respectively.}
    \label{fig:r0-xi-g}
\end{figure}

Fig.~\ref{fig:r0-xi-L} illustrates that a lower geometric index results in a more spacious allowed region. For $\sigma_0=4$ (upper panels), the contraction of allowed region occurs smoothly, aligning with the gradual decrease in outflow speed as ${\sf g}$ increases. While for $\sigma_0=50$ (bottom panels), with which the outflow speed as a function of ${\sf g}$ behaves like a step function, the allowed regions are almost equally spacious when ${\sf g}$ is not so closed to 1. When ${\sf g}=0.81$, as shown in the bottom panels, the allowed regions shrink suddenly, corresponding to the sudden drop of outflow speed.  It is also worth noticing here that some regions cannot be seen, such as the cases of ${\sf g}=0.64$ and ${\sf g}=0.81$ on the left panel of Fig.~\ref{fig:r0-xi-L}, because the energy extraction is not allowed under those sets of parameters \footnote{Situations are similar in Fig.~\ref{fig:r0-xi-g}. We will not mention it again.}. Obviously, there is a threshold of geometric index for energy extraction with determined black hole spin, local magnetization and streamline of bulk plasma. It would be figured out more easily by calculating the covering factor for energy extraciton, which will be discussed in Sect.~\ref{sec:chi}.

\begin{figure}
    \centering
    \hspace{-2.5mm}
    \includegraphics[width=0.48\textwidth]{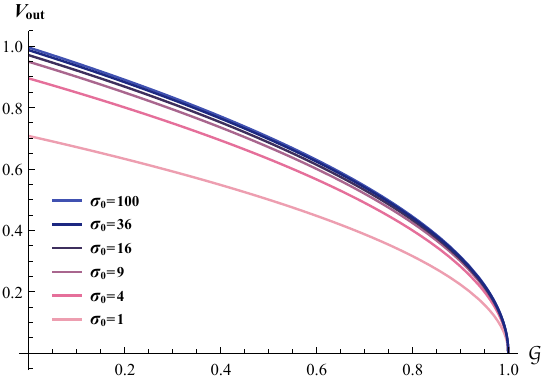}
    \hspace{2.5mm}
    \includegraphics[width=0.48\textwidth]{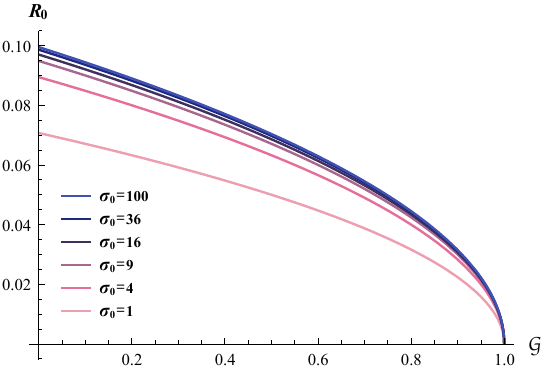}
    \hspace{-2.5mm}
    \caption{Outflow speed (left panel) and reconnection rate (right panel) as functions of the guide field fraction in quasi-3D model.}
    \label{fig:vout-R0-g}
\end{figure}

In Fig.~\ref{fig:vout-R0-g}, we plot outflow speed and local reconnection rate predicted in the quasi-3D model as functions of guide field fraction for multiple values of local magnetization. Both the outflow speed and local reconnection rate decrease monotonically with increasing ${\cal G}$. One can see from Eq.~\eqref{eq:R0g} that the dependence of $R_0$ on ${\cal G}$ tends to be $\sqrt{1-{\cal G}}$ when $\sigma_0\gg 1$ and that is why two panels in Fig.~\ref{fig:vout-R0-g} appear similar.

\begin{figure}
    \centering
    \includegraphics[width=0.45\textwidth]{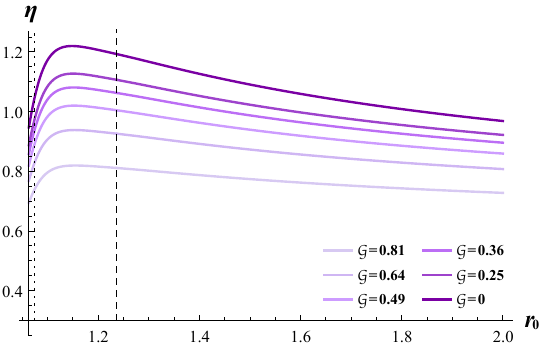}
    \hspace{2.5mm}
    \includegraphics[width=0.45\textwidth]{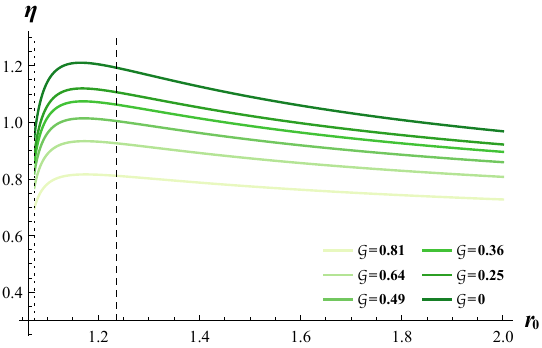}
    \hspace{-2.5mm}
    \includegraphics[width=0.45\textwidth]{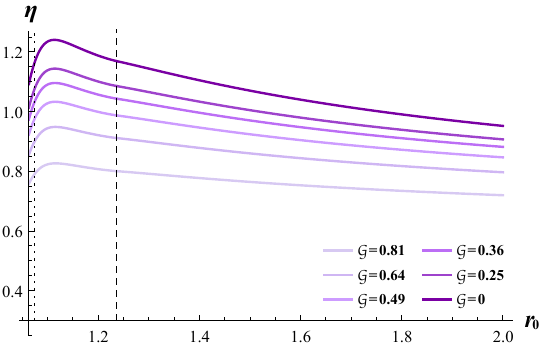}
    \hspace{2.5mm}
    \includegraphics[width=0.45\textwidth]{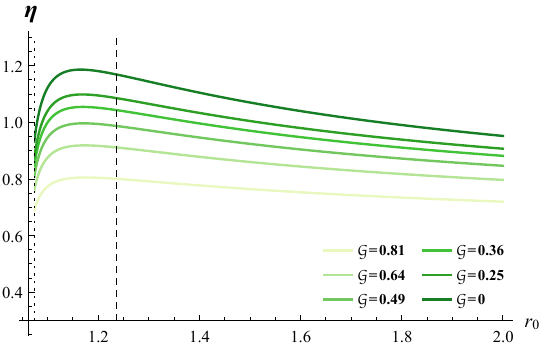}
    \hspace{-2.5mm}
    \includegraphics[width=0.45\textwidth]{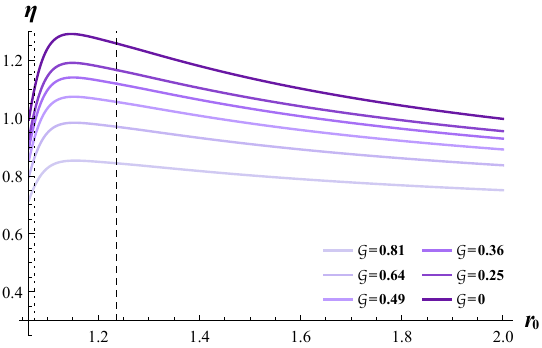}
    \hspace{2.5mm}
    \includegraphics[width=0.45\textwidth]{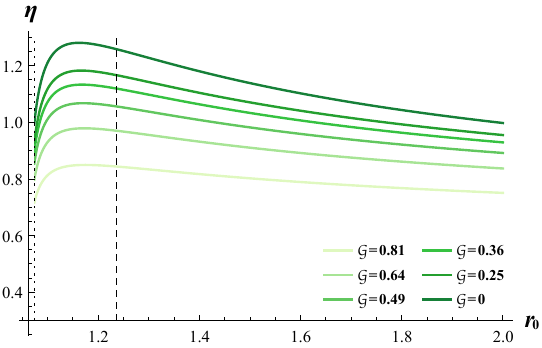}
    \hspace{-2.5mm}
    \includegraphics[width=0.45\textwidth]{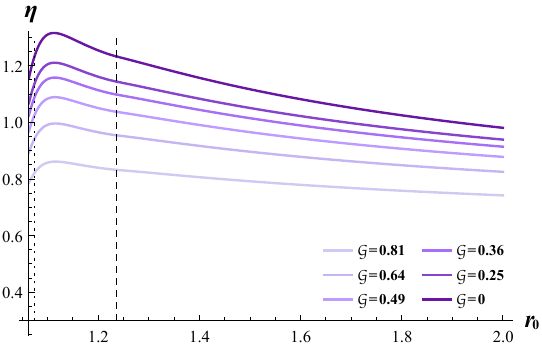}
    \hspace{2.5mm}
    \includegraphics[width=0.45\textwidth]{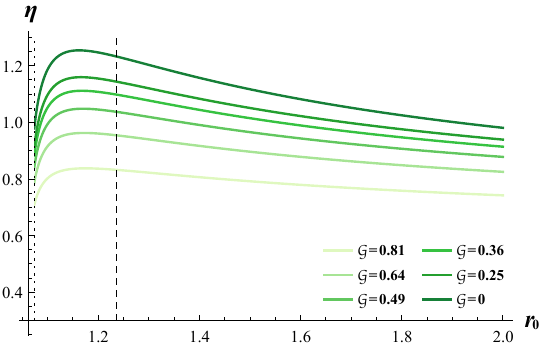}
    \caption{Efficiencies of energy extraction for multiple values of guide field fraction in quasi-3D model with $a=0.998$, $\xi_B=0$, $\sigma_0=4$ (panels on first and second rows) and $\sigma_0=50$ (panels on third and fourth rows), $\xi_B=0$ (panals on first and third rows) and $\xi_B=\pi/12$ (panels on second and fourth rows), combined streamline (left panels) and circular streamline (right panels). The dotted and dashed lines represent the photon sphere and ISCO, respectively.}
    \label{fig:eta-g}
\end{figure}

We present the efficiencies of energy extraction as functions of $r_0$ for multiple values of guide field fraction in Fig.~\ref{fig:eta-g}, with $a=0.998$, $\xi_B=0$ and $\xi_B=\pi/12$, $\sigma_0=4$ and $\sigma_0=50$, combined and circular streamlines, respectively. Besides, the allowed regions for energy extraction in $\xi_B-r_0$ planes for multiple values of guide field fraction are also plotted in Fig.~\ref{fig:r0-xi-g}. The efficiencies of and the allowed regions for energy extraction with two kinds of streamlines depend on the orientation angle in similar ways which were argued when discussing the results from the anti-parallel model above and the results in Ref.~\cite{Work1}. While distinct from the results based on the anti-parallel model, the decreases in efficiencies and shrinkages of allowed regions with increasing guide field fraction exhibit consistent trends across different values of local magnetization, corresponding to the similar trends in outflow speed as a function of ${\cal G}$ for multiple values of $\sigma_0$.

\subsection{Covering factor for energy extraction}
\label{sec:chi}

According to the discussions in Ref.~\cite{Work1}, we define the value
\begin{equation}
    \chi=\frac{S_{\rm EE}}{\pi\left(r_{\rm ergo}-r_{\rm EH}\right)}
    \label{eq:chi}
\end{equation}
to be the covering factor for energy extraction from a rotating black hole via magnetic reconnection, where $S_{\rm EE}$ is the "area" of the allowed region for energy extraction in $\xi_B-r_0$ plane. It depends on the black hole spin, local magnetization, streamline of bulk plasma and also the parameters in the reconnection model, as discussed in this work. This quantity reflects how probable the energy extraction could be realized via magnetic reconnection occurring within the ergo region of a rotating black hole, by assuming the occurrence of magnetic reconnection in any direction on any radius of the equatorial plane within the ergo-sphere is equally probable. This assumption is actually unrealistic in general. As we know, the local direction along which the magnetic reconnection occurs is highly dependent on the large scale configuration of magnetic field. Moreover, it is idealistic to presume that the magnetization of accretion flow in the ergo region distributes uniformly. Naturally, in the real accretion flow magnetic reconnections would never occur equally probably in different positions. However, for a primitive trial, it is not bad for us to analyze the covering factor to judge the capability of an accretion system on extracting energy from the central black hole via magnetic reconnection.

%This value reflects how probable the energy extraction could succeed via magnetic reconnection occurring in bulk plasma within ergosphere of a rotating black hole, by assuming that the magnetic reconnection is equally probable to occur along any direction on any position on the equatorial plane within ergosphere. It depends on the black hole spin, local magnetization, streamline of bulk plasma and also the parameters in the reconnection model, as discussed in this work. From another perspective, this value quantifies the capability of an accretion system—characterized by black hole spin, magnetization and the streamline of accretion flow—in extracting energy from the central black hole via magnetic reconnection. In Ref.~\cite{Work1}, we figured out that plunging bulk plasma is more capable of extracting energy than circularly flowing plasma by comparing values of $\chi$. Moreover, solving $\chi=0$ helps to determine the minimally allowed values of black hole spin and local magnetization for successful energy extraction

\begin{figure}
    \centering
    \includegraphics[width=0.48\textwidth]{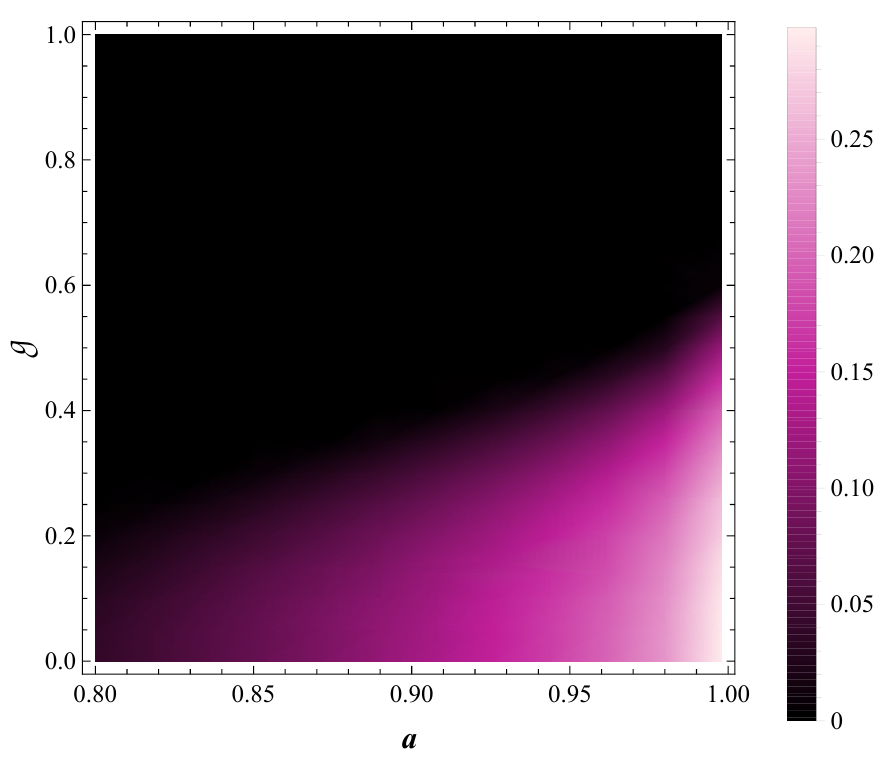}
    \hspace{3mm}
    \includegraphics[width=0.48\textwidth]{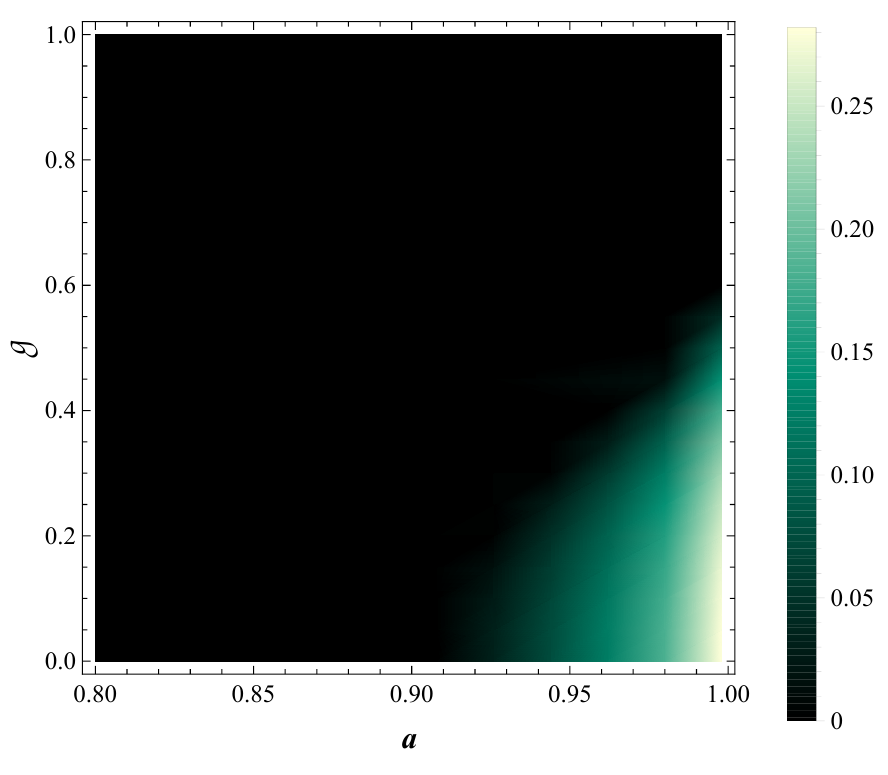}
    \hspace{-3mm}
    \includegraphics[width=0.48\textwidth]{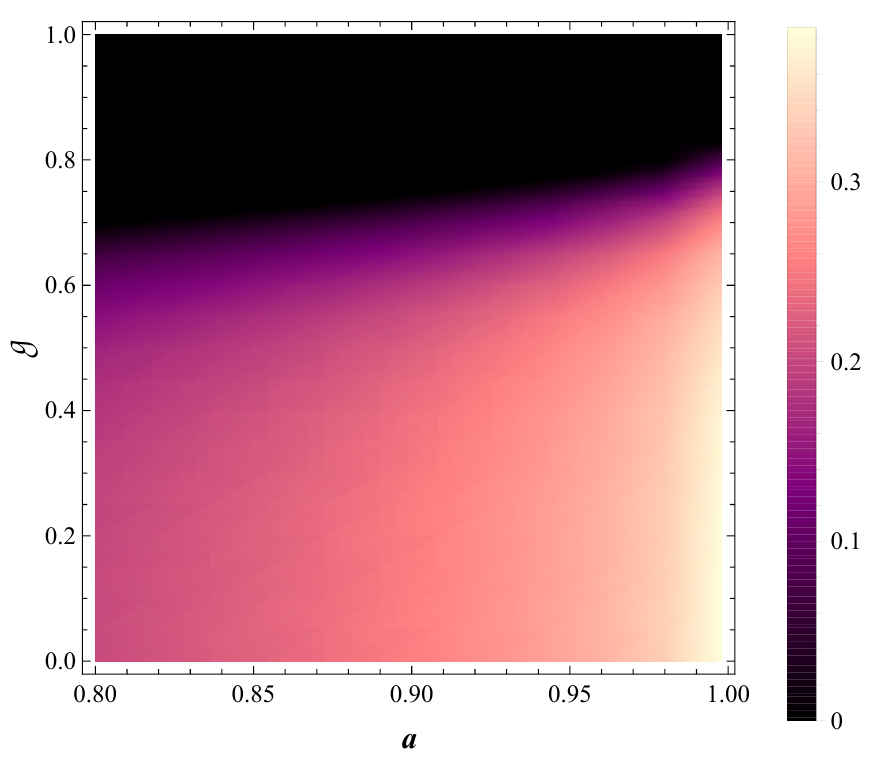}
    \hspace{3mm}
    \includegraphics[width=0.48\textwidth]{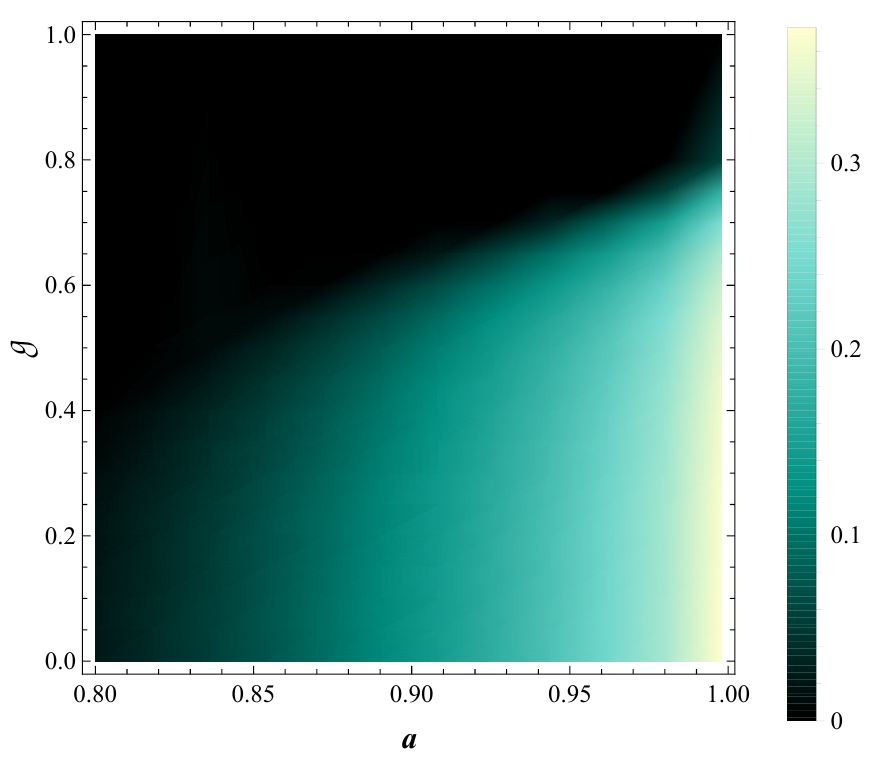}
    \caption{Distributions of covering factor, whose values are represented by lightness of colors, in $a-{\sf g}$ planes for the anti-parallel model, with $\sigma_0=4$ (top panels) and $\sigma_0=50$ (bottom panals), combined (left panels) and circular (right panels) streamlines.}
    \label{fig:chi-ag-L}
\end{figure}

\begin{figure}
    \centering
    \includegraphics[width=0.48\textwidth]{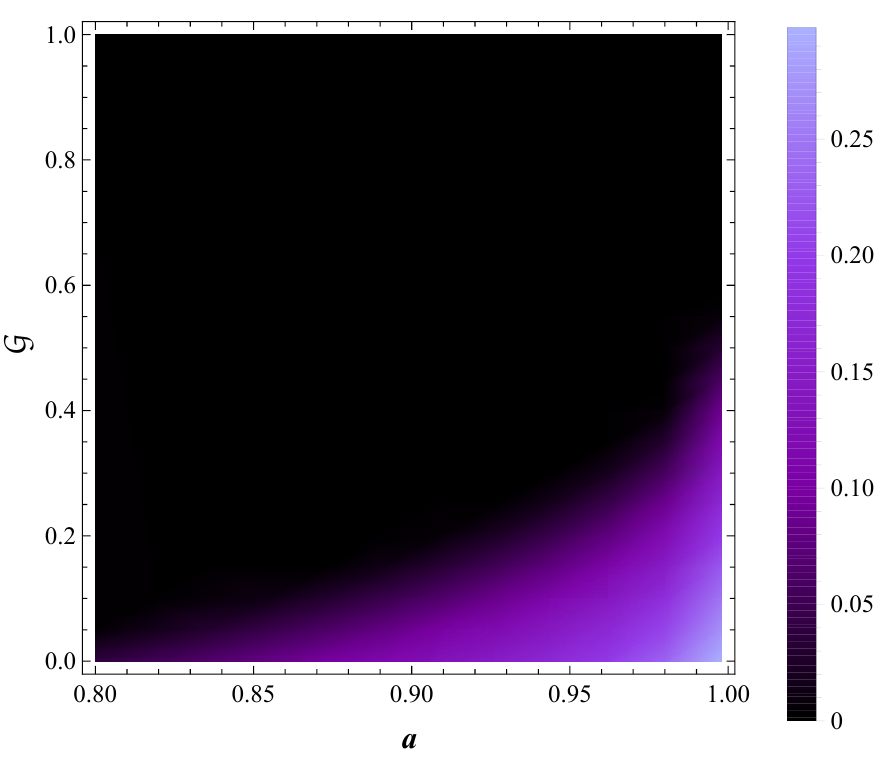}
    \hspace{3mm}
    \includegraphics[width=0.48\textwidth]{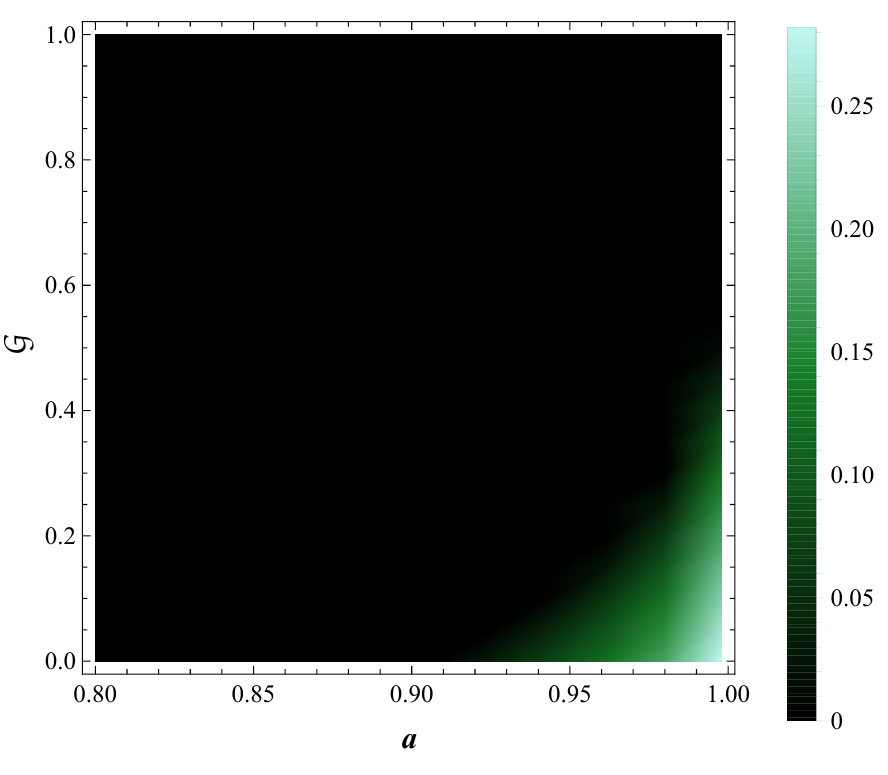}
    \hspace{-3mm}
    \includegraphics[width=0.48\textwidth]{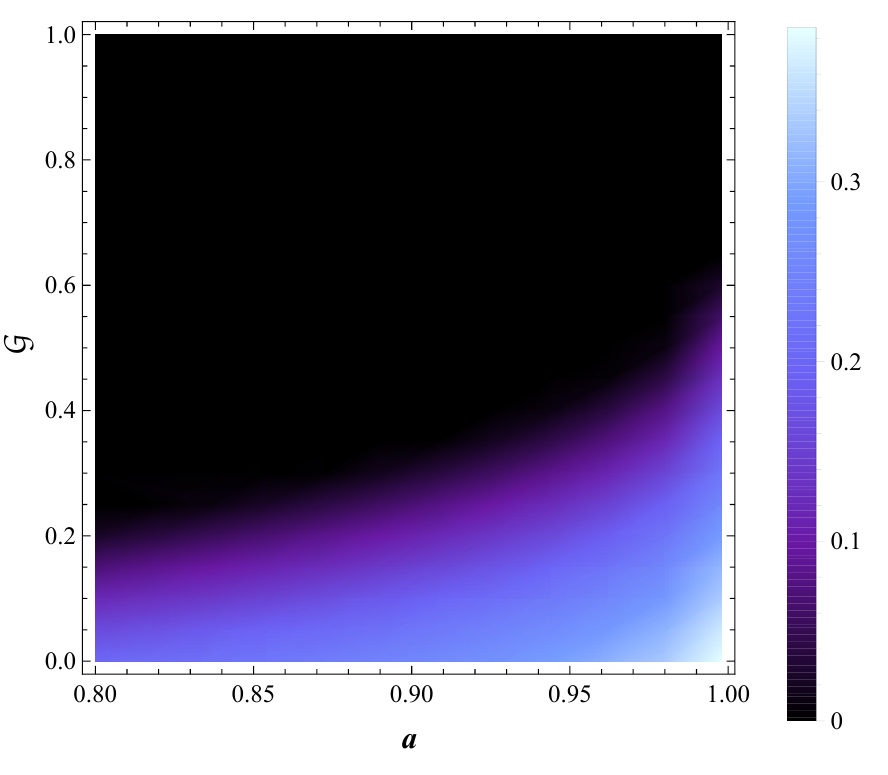}
    \hspace{3mm}
    \includegraphics[width=0.48\textwidth]{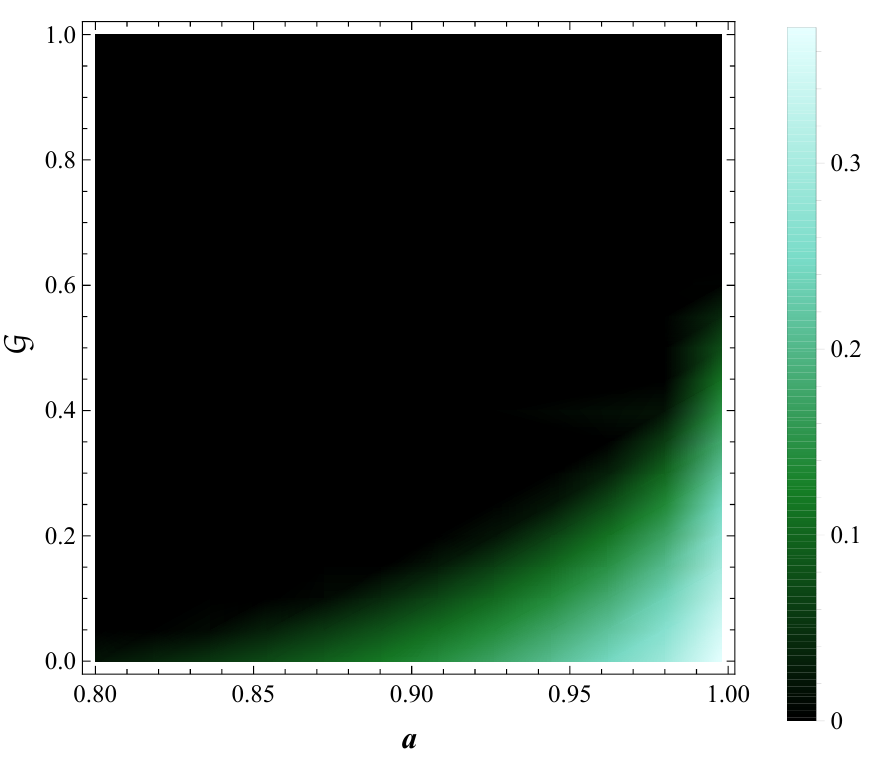}
    \caption{Distributions of covering factor, whose values are represented by lightness of colors, in $a-{\cal G}$ planes for quasi-3D model, with $\sigma_0=4$ (top panels) and $\sigma_0=50$ (bottom panals), combined (left panels) and circular (right panels) streamlines.}
    \label{fig:chi-ag-g}
\end{figure}

In Fig.~\ref{fig:chi-ag-L}, we plot the distributions of $\chi$ in $a-{\sf g}$ planes, assuming that the magnetic reconnection processes occur based on the description of the anti-parallel model, with $\sigma_0=4$ and $\sigma_0=50$, combined and circular streamlines. Here, the lightness of colors represent the values of $\chi$. Similarly, the distributions of $\chi$ in $a-{\cal G}$ planes are plotted in Fig.~\ref{fig:chi-ag-g}. The black areas on these figures represent the regions where energy extraction could never succeed. It is convenient to visibly determine the minimally allowed values of geometric index or guide field fraction for energy extraction from these plots with a given black hole spin. For example, from the bottom left panel in Fig.~\ref{fig:chi-ag-L}, one can figure out it is required that ${\sf g}\lesssim 0.65$ when $a\simeq 0.8$ while ${\sf g}\lesssim 0.8$ when $a\simeq 1$ for successful energy extraction via magnetic reconnection occurring based on the prediction of the anti-parallel model, when the bulk plasma plunges with the local magnetization $\sigma_0=50$. Similarly, from the top right panel in Fig.~\ref{fig:chi-ag-g}, one can see that energy extraction via magnetic reconnection, as predicted by the quasi-3D model, is impossible for $a\simeq 0.8$ while it is required that ${\cal G}\lesssim 0.5$ for $a\simeq 1$, when the bulk plasma flows circularly with $\sigma_0=4$. 

Furthermore, comparisons between the resultant covering factors indicate that the plunging bulk plasma is more capable of extracting energy than the circularly flowing one, especially when the black hole spin is not extreme such that the plunging region is spacious. However, the strength of the plunging bulk plasma diminishes with the shrinkage of the plunging region. According to the results in Ref.~\cite{Work1} where the anti-parallel model was adopted with ${\sf g}=0.49$, $a\gtrsim 0.8$ is generally required if the bulk plasma always flows circularly. While the requirement of $a\gtrsim 0.8$ for energy extraction in the circularly flowing bulk plasma can be extended to most values of ${\sf g}$, as shown in the right panels of Fig.~\ref{fig:chi-ag-L} . One can also see from the right panel of Fig.~\ref{fig:chi-ag-g} that $a\gtrsim 0.9$ and $a\gtrsim 0.8$ are required for successful energy extraction in the circularly flowing bulk plasma with $\sigma_0=4$ and $\sigma_0=50$ respectively, for an arbitrary value of ${\cal G}$ when the quasi-3D model is adopted.

Additionally, Fig.~\ref{fig:chi-ag-g} indicates that the values of $\chi$ in $a-{\cal G}$ planes vary in a similar way for whatever low or high local magnetization. It is consistent with the variations of outflow speed, whose dependences on the guide field fraction and local magnetization are separable. The situation changes if the anti-parallel model is adopted. In the $a-{\sf g}$ planes, the shape of the allowed region for energy extraction (region with $\chi>0$) gradually approaches a rectangle as local magnetization increases (compare the top and bottom panels in Fig.~\ref{fig:chi-ag-L}). It corresponds to the outflow speed as a function of geometric index, which approaches a step function when $\sigma_0\rightarrow \infty$

\begin{figure}
    \centering
    \hspace{-2mm}
    \includegraphics[width=0.48\textwidth]{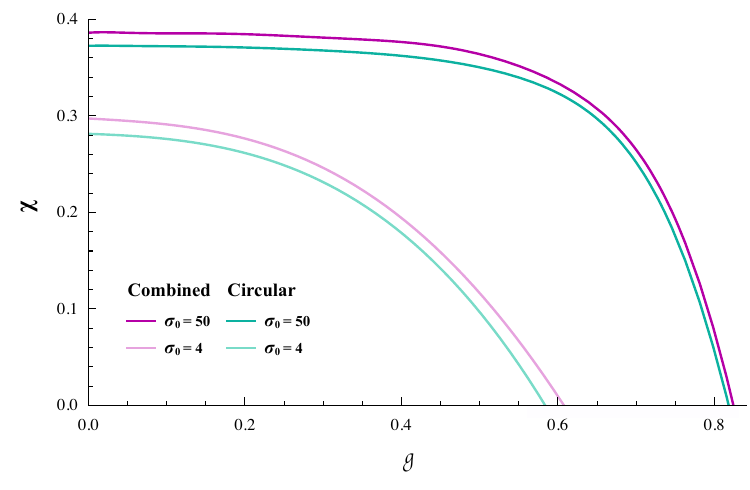}
    \hspace{2mm}
    \includegraphics[width=0.48\textwidth]{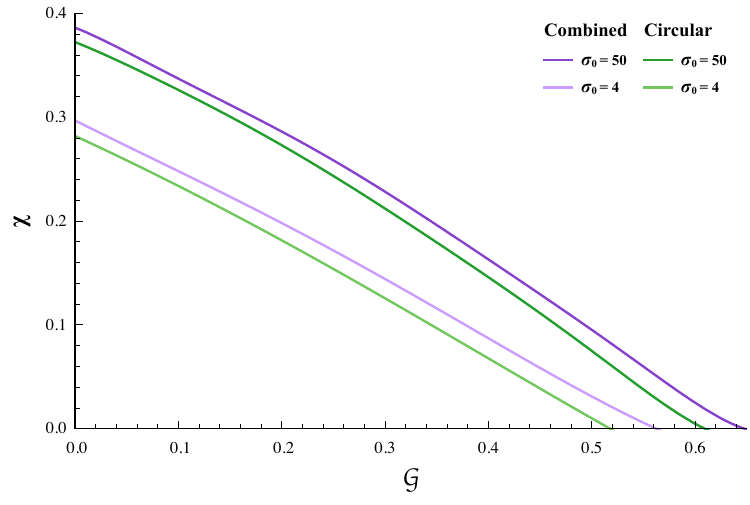}
    \hspace{-2mm}
    \caption{Covering factors as functions of geometric index (left panel) and guide field fraction (right panel), with $a=0.998$, $\sigma_0=4$ (lightly colored) and $\sigma_0=50$ (deeply colored), combined (purple lines) and circular (green lines) streamlines.}
    \label{fig:chi-g}
\end{figure}

We also plot $\chi$ as functions of either ${\sf g}$ or ${\cal G}$ in Fig.~\ref{fig:chi-g}, with $a=0.998$, $\sigma_0=4$ and $\sigma_0=50$, combined and circular streamlines, respectively. The strength of the plunging bulk plasma almost vanishes since the black hole spin is nearly extreme. It is obvious that $\chi$ as functions of ${\sf g}$ are akin to step functions when the local magnetization is high. While the variations of $\chi$ as functions of ${\cal G}$ remain smooth regardless of whether high or low magnetizations.

\subsection{Averaged covering factor for energy extraction}
\label{sec:ave-chi}

\begin{figure}
    \centering
    \includegraphics[width=0.48\textwidth]{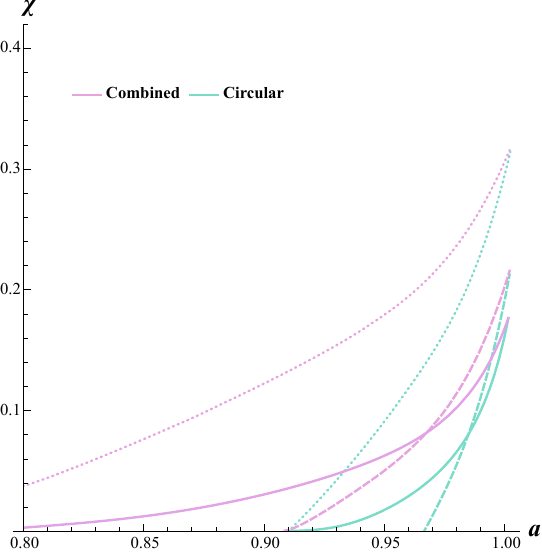}
    \includegraphics[width=0.48\textwidth]{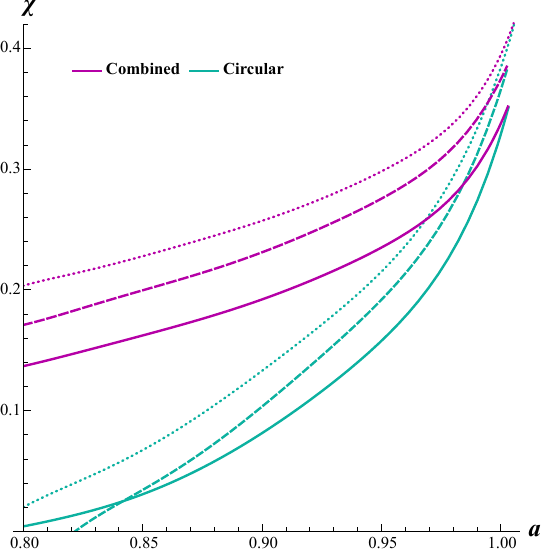}
    \caption{Solid lines represent averaged covering factors weighted by geometric index, with $a=0.998$, $\sigma_0=4$ (left panel) and $\sigma_0=50$ (right panel), combined (plum lines) and circular (turquoise lines) streamlines. Dotted and dashed lines represent the covering factors with ${\sf g}=0$ and the ${\sf g}$ maximizing the reconnection rate, respectively.}
    \label{fig:chi-ave-L}
\end{figure}

\begin{figure}
    \centering
    \includegraphics[width=0.48\textwidth]{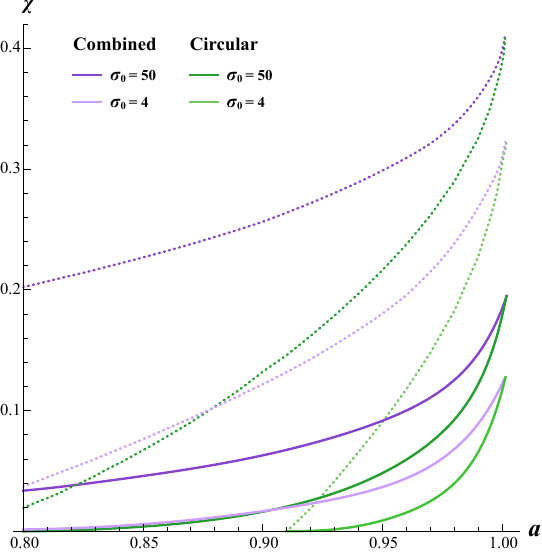}
    \caption{Solid lines represent averaged covering factors weighted by geometric index, with $a=0.998$, $\sigma_0=4$ (lightly colored lines) and $\sigma_0=50$ (deeply colored lines), combined (purple lines) and circular streamlines (green lines). Dotted lines represent the covering factors with ${\cal G}=0$.}
    \label{fig:chi-ave-g}
\end{figure}

\begin{figure}
    \centering
    \includegraphics[width=0.48\textwidth]{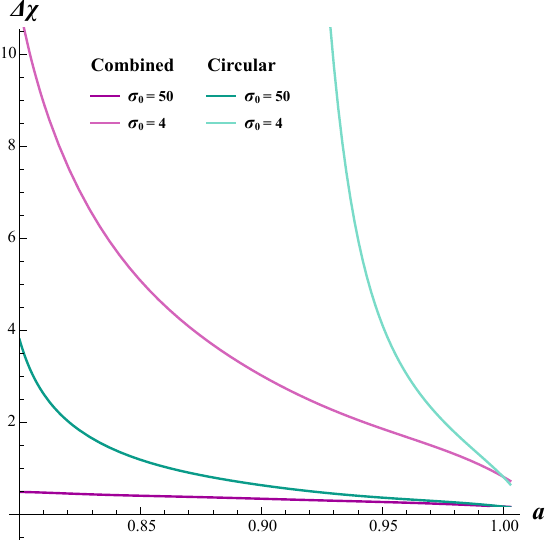}
    \hspace{2mm}
    \includegraphics[width=0.48\textwidth]{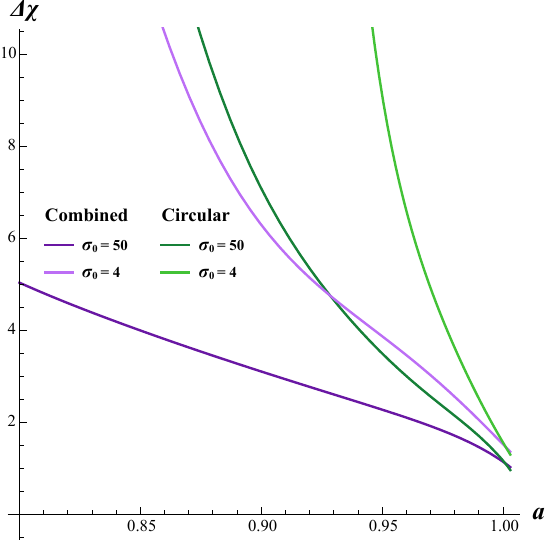}
    \caption{The deviation between covering factor with $G=0$ and the averaged covering factor weighted by reconnection rate, when the anti-parallel model (left panel) and the quasi-3D model (right panel) are adopted.}
    \label{fig:dev}
\end{figure}

In astrophysical scenario, magnetic reconnection processes occur stochastically and occasionally explosively \cite{Jia:2023iup}, during which the values of parameters should be random even if the occurrences of magnetic reconnection align with the descriptions of the reconnection models. From an observational standpoint, people would be more interested in the probability of energy extraction with the parameters in reconnection models integrated out. For this purpose, we recommend to calculate the averaged covering factor as:
\begin{equation}
    \left<\chi\right>_{w_0}=\frac{\int_0^1 w_0(G)\chi(G) dG}{\int_0^1 w_0(G) dG}
    \label{eq:ave_chi}
\end{equation}
where $w_0(G)$ is the weight dependent on parameters in different reconnection models respectively (represented by $G$ uniformly). The averaged covering factor may quantify an overall capability of an accretion system in extracting energy via magnetic reconnection which occurs stochastically with random values of parameters in reconnection models, if the weight is chosen suitably.

Let us review the properties of reconnection rate. It is defined to be the local Mach number on the reconnection point, which is primarily used to determine whether the magnetofluid could be treated as an incompressible flow. This means that the lower the reconnection rate, the more incompressible the magnetofluid at the reconnection point will be. Subsequently, the magnetic reconnection process would be hard to succesfully occur if the reconnection rate is low. On the other hand, the reconnection rate is proportional to the inflow speed on the reconnection point. Hence a higher reconnection rate means that the magnetic reconnection process occurs faster. When the magnetic reconnection processes happen multiple times stochastically in the ergo region, it is reasonable to predict that the processes with higher reconnection rate would occur more times than the processes with lower reconnection rate. In this sense, we recommend to choose the reconnection rate to be the weight in Eq.~\eqref{eq:ave_chi}.

In Fig.~\ref{fig:chi-ave-L}, we plot the averaged covering factor weighted by reconnection rate, adopting the anti-parallel model, as functions of black hole spin by solid lines, with $\sigma_0=4$ (left panel) and $\sigma_0=50$ (right panel), combined and circular streamlines. The $G$ in Eq.~\eqref{eq:ave_chi} represents the geometric index now. Meanwhile, the covering factors with ${\sf g}=0$ (Sweet-Parker approach) and with the ${\sf g}$ maximizing reconnection rate are plotted by dotted and dashed lines, respectively. In a similar manner, the averaged covering factors weighted by reconnection rate, under the quasi-3D model, are displayed in in Fig.~\ref{fig:chi-ave-g}. The covering factors for ${\cal G}=0$, which also maximizes the reconnection rate, are shown together by dotted lines.

When the anti-parallel model is adopted, the covering factor would definitely become higher when ${\sf g}\rightarrow 0$. However, the infinitesimal reconnection rate indicates that the likelihood for the occurrence of magnetic reconnection with ${\sf g}\simeq 0$ is nearly nonexistent. The approximation of the null geometric index made in previous works \cite{CA2021} may overestimate the capability of an accretion system in energy extraction via magnetic reconnection. In Fig.~\ref{fig:dev}, We plot the deviation between covering factor with $G=0$ (denoted by $\chi_0$) and the averaged one as a function of black hole spin, defined as
\begin{equation}
    \Delta\chi=\frac{\chi_0-\left<\chi\right>_{R_0}}{\left<\chi\right>_{R_0}}
    \label{eq:dev}
\end{equation}
This value indicates how the energy extraction capability is overestimated when relying solely on the Sweet-Parker configuration for the magnetic reconnection process. The left panel depicts the deviation in the case of the anti-parallel model. The overestimation when $\sigma_0=4$ is severe, especially when the black hole spin is low. Opting ${\sf g}$ that maximizes the reconnection rate would be better. While it is noteworthy that this approach underestimates the capability when the black hole spin is relatively low, because the cases with ${\sf g}\simeq 0$ are completely ignored. The deviation between covering factor with fixed ${\sf g}$ and the averaged one become less significant when the local magnetization gets larger, as indicated by the right panel of Fig.~\ref{fig:chi-ave-L} and the left panel of Fig.~\ref{fig:dev}. We mentioned in Sect.~\ref{sec:para} that the outflow speed as a function of geometric index approaches a step function when $\sigma_0\rightarrow \infty$. Therefore, situations of energy extraction via magnetic reconnection are similar for all the allowed values of ${\sf g}$ unless ${\sf g}\rightarrow 1$. It can be anticipated that the covering factor with certain ${\sf g}$ and the averaged covering factor, both as functions of black hole spin, get closer and closer when the local magnetization approaches infinity. Thus, it is acceptable to pick an arbitrary value of ${\sf g}$ when assessing the capability of an accretion system in energy extraction via magnetic reconnection when the local magnetization is high.

Based on the results of the quasi-3D model, similarly, the covering factor when ${\cal G}\simeq 0$ would be higher than the averaged covering factor. The case of ${\cal G}=0$ yields the highest reconnection rate at the meantime. However, since the reconnection rate decreases slowly when the guide field fraction is increased, the deviation $\Delta\chi$ is still noticeable. What is worse, in the quasi-3D model, the trend of outflow speed as a function of guide field fraction are independent of the local magnetization. Increasing the magnetization would not considerably remit the overestimation, as indicated in Fig.~\ref{fig:chi-ave-g} and the right panel of Fig.~\ref{fig:dev}.

\section{Summary and discussions}
\label{sec:sum}

In this work, we extend our discussions in Ref.~\cite{Work1}, regarding energy extraction via the Comisso-Asenjo process from a rotating black hole, to the situations of unfixed parameters in reconnection models. Both the anti-parallel model, an analytical reconnection model under MHD scheme, and the quasi-3D model, the model based on the results of numerical simulations by PIC method, are employed to describe the magnetic reconnection process. 

We present the efficiencies of energy extraction for multiple values of geometric index (when the anti-parallel model is adopted) or guide field fraction (when the quasi-3D model is adopted). Meanwhile, the allowed regions for energy extraction in $\xi_B-r_0$ planes for multiple values of geometric index or guide field fraction are exhibited. In line with the discussions in Ref.~\cite{Work1}, we compare the situations involving plunging bulk plasma and circularly flowing bulk plasma to ensure the results in this work are consistent with those in Ref.~\cite{Work1}. We conclude that a higher geometric index or a higher guide field fraction generate lower outflow speed of plasmoids, making the energy extraction via magnetic reconnection harder.

We present the distributions of covering factor, introduced in Ref.~\cite{Work1}, in $a-{\sf g}$ and $a-{\cal G}$ planes for some representative values of local magnetization, from which it is convenient to determine the maximally allowed geometric index or guide field fraction for a given black hole spin, or conversely, to determine the minimally allowed black hole spin for a given geometric index or guide field fraction. In order to quantify the overall capability of an accretion system in extracting energy via magnetic reconnection, we recommend to calculate the averaged covering factor weighted by reconnection rate. It is evident that selecting the Sweet-Parker configuration (for ${\sf g}\rightarrow 0$ or ${\cal G}\rightarrow 0$) merely would overestimate the capability of an accretion system in energy extraction via magnetic reconnection. We quantify this overestimation by measuring the deviation between the covering factor for ${\sf g}\simeq 0$ or ${\cal G}\simeq 0$ and the averaged covering factor weighted by reconnection rate. We figure out that, when the anti-parallel model is adopted, the overestimation of covering factor is mitigated in the limit of high magnetization. Thus, it is acceptable if one just opts ${\sf g}\simeq 0$ to consider the energy extraction via magnetic reconnection when the magnetization is extremely high. However, when the quasi-3D model is under consideration, the overestimation remains severe even if the local magnetization is extremely high. 

It is worth noticing that the anti-parallel model proposed in Ref.~\cite{Liu2017} is incomplete, since the geometric index is induced as a phenomenological parameter. There must be essential factors controlling the value of the geometric index when the magnetic reconnection occurs in bulk plasma, such as the plasma pressure gradient \cite{2011PhPl...18k1204W,2014PhPl...21b2113L} or temperature anisotropies upstream \cite{2013PhPl...20f1201E}. One may establish the relations between the energy extraction via magnetic reconnection and some physical quantities of plasma, which are reflected by the value of geometric index, after improving the anti-parallel model.

Meanwhile, the quasi-3D model is still not trustworthy, which expects the outflow speed and the reconnection rate in oversimplified ways. What is worse, question of whether the configuration of magnetic field described by Eq.~\eqref{eq:iniB} at the local scale could naturally form remains in doubt. Further researches, either analytical or numerical, on three-dimensional models are needed to help us elucidate how the magnetic field lines reconnect in a real astrophysical system.

Furthermore, this work is merely a preliminary trial, considering only two simple reconnection models separately. In real astrophysical systems, magnetic reconnection processes may occur in numerous ways simultaneously. For example, it is likely that the magnetic reconnections obeying the description of the anti-parallel model and the quasi-3D model occur within a single accretion system. In this case, the overall capability of the accretion system in energy extraction, which could be quantified by the averaged covering factor weighted by reconnection rate or other physical quantities, should then be discussed more meticulously.

Last but not least, since the magnetic reconnection is assumed to occur near the event horizon in this work, it is necessary to analyze the gravitational effects on the properties of magnetic reconnection. Some works indicated that the properties of outflow speed and reconnection rate would be modified significantly by spacetime curvature \cite{CA2017,Comisso:2018ark,Fan:2024fcy}. However, Ref.~\cite{Shen:2024plw} argued that the gravitational effect should be negligible at the local scale. Further exploration is needed to better understand the gravitational effect on the magnetic reconnection process.

\section*{Acknowledgement}

Ye Shen and Ho-Yun YuChih contribute equally to this work. We would like to thank Prof. Bin Chen for his support and suggestions. We thank the anonymous referee for valuable comments and constructive suggestions.

%%%%%%%%%%%%%%%%% APPENDICES %%%%%%%%%%%%%%%%%%%%%
\appendix

%\newpage
\bibliographystyle{utphys}
\bibliography{references}

\end{document}